\pdfoutput=1
\documentclass[reprint, amsmath, amssymb, aps]{revtex4-2}
\usepackage{color}
\usepackage[T1]{fontenc}
\usepackage{lmodern}
\usepackage{textcomp}
\usepackage{hyperref}
\hypersetup{colorlinks=false, hypertexnames=false}
\usepackage{graphicx}
\usepackage{dcolumn}
\usepackage{bm}

\begin{document}

\preprint{APS/123-QED}


\title{Ultra-wideband electrically-tuned mid-infrared on-chip parametric oscillator}

\author{Alexander Y. Hwang\textsuperscript{1}}%
\author{Hubert S. Stokowski\textsuperscript{1}}%
\author{Luke Qi\textsuperscript{1}}%
\author{David K. Concepcion\textsuperscript{1}}%
\author{Geun Ho Ahn\textsuperscript{1}}%
\author{Ethan Rosenfeld\textsuperscript{1}}%
\author{Taewon Park\textsuperscript{1}}%
\author{Devin J. Dean\textsuperscript{1}}
\author{Martin M. Fejer\textsuperscript{1}}%
\author{Amir H. Safavi-Naeini\textsuperscript{1,*}}%

\affiliation{%
 \textsuperscript{1}E.L. Ginzton Laboratory, Stanford University, Stanford, CA, 94305, USA
}

\begin{abstract}

Developing compact, broadband mid-infrared coherent sources for applications in spectroscopy and sensing remains a pressing challenge in photonics. However, material limitations and integration constraints have restricted the accessible wavelengths and operation bandwidths of current mid-infrared lasers. Here, we address these challenges by developing a nonlinear integrated photonic device that converts a fixed-wavelength near-infrared pump laser into broadly tunable mid-infrared light. Our device, an optical parametric oscillator (OPO) integrated on thin-film lithium niobate, generates 22 THz of multi-milliwatt, voltage-tunable radiation from 2.7-3.4 \textmu m, a region typically difficult to access but vital for environmental, chemical, and biological sensing. By introducing an on-chip-tunable OPO architecture taking advantage of the Vernier effect, we obtain electrical control of the emission wavelengths from coarse, multi-THz scales down to continuous, sub-100-GHz mode-hop-free tuning ranges. This work establishes a robust platform for a new class of compact, widely tunable mid-infrared sources with potential for future scaling.

\end{abstract}

\maketitle


\section{Introduction}

For decades, mid-infrared laser technologies have attracted significant research interest owing to their broad applications in chemical sensing \cite{chikkaraddySinglemoleculeMidinfraredSpectroscopy2023}, biosensing \cite{liangModulatedRingdownComb2025}, environmental monitoring \cite{ycasHighcoherenceMidinfraredDualcomb2018a, longNanosecondTimeresolvedDualcomb2024}, communications \cite{zouHighcapacityFreespaceOptical2022}, and defense. Recently, the performance of these laser sources has advanced tremendously \cite{kazakovDrivenBrightSolitons2025, taschlerFemtosecondPulsesMidinfrared2021, fuchsbergerContinuouslyWidelyTunable2025, zhouHighPerformanceMonolithic2017, yuSiliconchipbasedMidinfraredDualcomb2018a}, expanding new commercial applications. Currently, one of the pressing challenges of this field lies in producing coherent mid-infrared sources that combine broadband wavelength operation with compact, scalable integration. Such sources would dramatically aid our ability to acquire dense, multidimensional spectral information in practical settings, but realizing them remains a formidable challenge.

Meanwhile, near-infrared integrated tunable lasers, especially from $1.2$-$1.6$ \textmu m, have made enormous progress \cite{siddharthUltrafastTunablePhotonicintegrated2025, zhouProspectsApplicationsOnchip2023, heimHybridIntegratedUltralow2025, luEmergingIntegratedLaser2024, tranExtendingSpectrumFully2022a}. These compact lasers can be rapidly tuned over broad ranges using low-power integrated electrical signals. Such advancements have been driven by combining III-V laser gain with mass-manufacturable silicon and silicon nitride photonics. However, these devices remain fundamentally limited to spectral windows where material gain is available.

Integrating III-V sources with nonlinear integrated photonics is a compelling strategy to operate in important but difficult-to-access spectral ranges such as the mid-infrared. Nonlinear photonic platforms, such as thin-film lithium niobate \cite{boesLithiumNiobatePhotonics2023, zhuIntegratedPhotonicsThinfilm2021a} and silicon nitride \cite{duttNonlinearQuantumPhotonics2024, nitissOpticallyReconfigurableQuasiphasematching2022, jiMultimodalityIntegratedMicroresonators2024}, have been accelerated in development by telecom, datacom, sensing, and quantum computing applications.   

Integrated optical parametric oscillators (OPOs), in particular, have recently emerged as exemplary nonlinear devices converting near-infrared lasers to widely tunable output in difficult-to-access wavelength regions \cite{hwangMidinfraredSpectroscopyBroadly2023a, ledezmaOctavespanningTunableInfrared2023a, perezHighperformanceKerrMicroresonator2023, pidgaykoVoltagetunableOpticalParametric2023, luPhotonicIntegratedCircuit2026}. These devices close spectral gaps from the visible to the mid-infrared with milliwatt level thresholds and compact footprints. However, broad wavelength tunability in these devices currently requires widely-tunable external pump lasers, substantial on-chip temperature tuning, or operation across multiple discrete devices. These methods all introduce complexities and diminish the advantages of OPO integration, limiting their applicability in many contexts.

\begin{figure*}[t]
  \begin{center}
      \includegraphics[width=\textwidth]{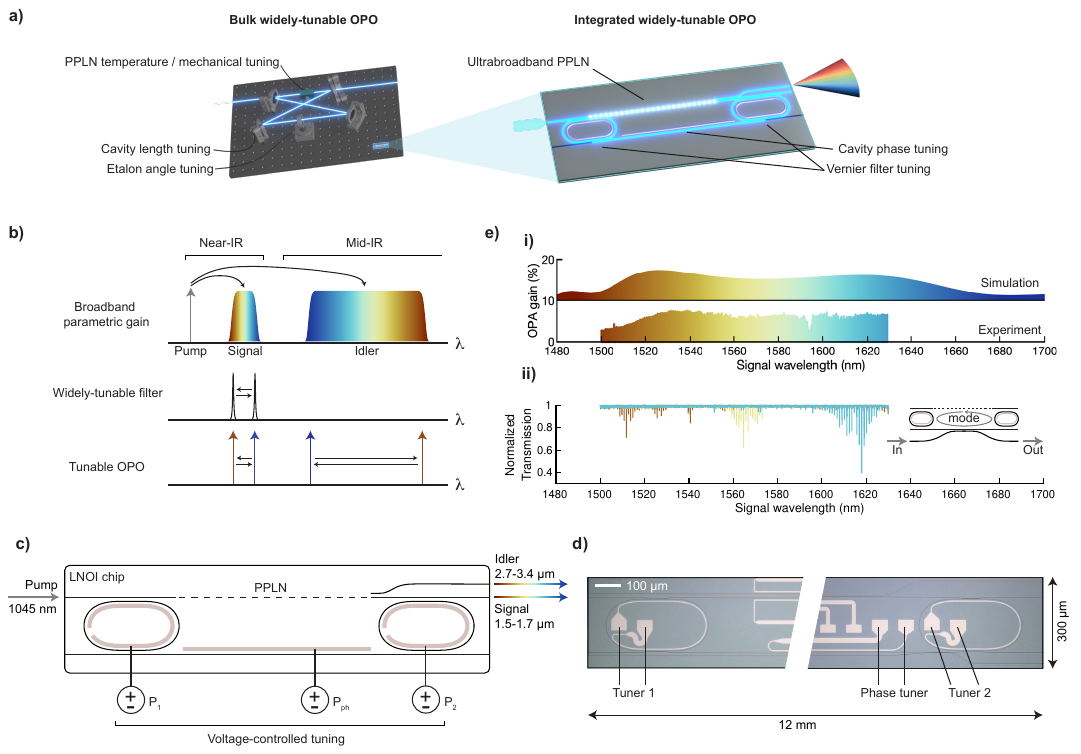}
  \end{center}
 \caption{\textbf{Vernier OPO concept and design.} a) Illustration comparing the mechanisms to achieve broad and precise tuning in a typical bulk OPO configuration versus the integrated OPO of this work. b) Strategy to achieve widely-tunable and highly-controlled mid-infrared oscillation. c) Photonic-chip-based realization, employing a voltage-controlled Vernier filter inside an optical parametric oscillator. d) Optical images of Vernier tuner cavities on the fabricated devices. e) Characterization of device gain and tunable Vernier modes. e.i) Simulated and experimentally-measured broadband OPA signal gain for a fixed pump wavelength of $1050$ nm with $77$ mW on-chip power. The experimental data is limited by our tunable probe laser's scan range. e.ii) Vernier modes tuned over three different heater voltages, seen as transmission dips when probing the OPO cavity from an external waveguide.
}
 \label{fig:fig1}
\end{figure*}

Here, we demonstrate an integrated OPO device architecture that provides tuning over broad ranges of mid-infrared wavelengths \textit{solely with on-chip electrical tuning}. The device unlocks wide tuning in the $3$ \textmu m wavelength region that is typically difficult to access but application-rich for gas/liquid sensing, communications, and breath sensing. Inspired by the successful device paradigms of near-infrared tunable lasers, we realize ultra-wideband tunability by introducing a design incorporating the Vernier effect within the OPO cavity. This enables a single device at constant temperature, pumped with a fixed $1045$ nm laser, to tune over $190$~nm of near-infrared and $660$~nm of mid-infrared with integrated electrical control. Our device successfully integrates much of the functionality and precision of broadly-tunable, mechanically-modulated bulk OPOs onto an integrated chip (Fig. \ref{fig:fig1}a). This architecture, combined with moderate optical and electrical power improvements in future devices, enables a new class of tunable mid-infrared sources that can leverage the scaling potential of nonlinear photonics and near-infrared lasers.

\section{Results}

\subsection{Vernier OPO design}

Our device's functionality emerges from a conceptually simple design combining broadband gain and widely tunable wavelength selectivity (Fig. \ref{fig:fig1}b). A fixed-wavelength near-infrared pump around $1045$ nm provides the broad gain via a $\chi^{(2)}$ nonlinear interaction to signal wavelengths from $1500$-$1700$ nm and idler wavelengths from $2700$-$3400$ nm. Then, a tunable filter inside the cavity precisely selects the mode to oscillate. The large wavelength separation afforded by the $\chi^{(2)}$ interaction is crucial to this architecture for three reasons. First, it enables bright mid-infrared oscillation from pumping in the near-infrared, where diode lasers can be mass-produced. Second, it allows resonating only the signal wave; such singly resonant \cite{vainioSinglyResonantCw2008a,montesUltracoherentSignalOutput2004} OPO architectures crucially have the most straightforward and reproducible tuning mechanism. Third, it enables all the critical components, i.e. gain and tunable photonics, of this mid-infrared-emitting device to be implemented and characterized in the near-infrared where tunable lasers are already widely available.

We implemented the device with a thin-film lithium niobate (TFLN) nonlinear photonic integrated circuit (Fig. \ref{fig:fig1}c,d). TFLN was chosen for its strong nonlinearity, mid-infrared transparency, and demonstrated ability to support low-loss, high-functionality photonic structures \cite{yuIntegratedFemtosecondPulse2022a, fengIntegratedLithiumNiobate2024, huHighefficiencyBroadbandOnchip2022a, chilesMidinfraredIntegratedWaveguide2014}. A $9$-mm long periodically-poled lithium niobate (PPLN) section utilizes dispersion engineering \cite{mishraUltrabroadbandMidinfraredGeneration2022b} to support broadband parametric gain. Incorporating two racetrack resonators with slightly different free spectral ranges inside the OPO creates highly tunable wavelength selectivity via the Vernier effect, which is used extensively in near-infrared tunable laser devices \cite{mortonIntegratedCoherentTunable2022, liIntegratedPockelsLaser2022a, guoEbandWidelyTunable2023, komljenovicWidelyTunableNarrowLinewidth2015}. Because the two resonators have slightly different mode spacings, their resonances coincide only at specific wavelengths; a small shift of one resonator moves this coincidence point across a much larger wavelength range. This enables wide wavelength tuning from small, low-power electrical signals. The Vernier filter's transmission wavelength can be controlled by applying thermo-optic phase shifts to the two racetrack tuner cavities and a phase section in between. A directional coupler taps off the mid-infrared idler light generated in the PPLN to the chip output with high efficiency. 

We achieve broadband parametric gain by dispersion engineering the PPLN waveguide geometry. In particular, non-degenerate optical parametric amplification (OPA) bandwidth becomes maximal when signal-idler group velocity mismatch (GVM) and total signal and idler group velocity dispersion (GVD) are minimized \cite{jankowskiDispersionengineeredH2Nanophotonics2021b}. Achieving these dispersion parameters generally requires large film thicknesses (${>}600$ nm). In turn, a deep etch depth enables strong mode confinement while reducing slab mode leakage. The optimized waveguide geometry (Fig. S\ref{fig:opa}a-b) broadens the bandwidth by canceling the first-order phase mismatch variation versus wavelength (GVM), leaving only the quadratic contribution (GVD). Our measured OPA gain versus signal wavelength (Fig. S\ref{fig:opa}d) verifies this GVD-limited phase mismatch, evidenced as a double peaked transfer function. The experimental data matches simulated transfer functions that include a small thickness change varying quadratically along the waveguide length. As the pump tunes to $1045$-$1050$ nm, the two peaks merge together, providing $20$~THz gain bandwidth (Fig. \ref{fig:fig1}e.i) with a normalized gain of $68$\%/W (Fig. S\ref{fig:opa}e). 

We design the OPO's Vernier-tunable photonics trading off the filter's insertion loss, tuning range, and mode selectivity, then validate with tunable laser mode spectroscopy. First, we aim to minimize the Vernier filter's insertion loss to allow for OPO threshold at reasonable power levels. Strongly overcoupling (${>}10$x) the tuner cavities to their feedlines reduces the round-trip loss of the Vernier filter to ${\sim}25$\% (Fig. S\ref{fig:vernier_design}c). Second, we choose a small tuner cavity length offset of $\Delta L = 6$ \textmu m to extend the Vernier tuning range to $20$ THz (Fig. S\ref{fig:vernier_design}d), to take advantage of our full nonlinear gain bandwidth. The length offset cannot be made too small, otherwise the Vernier filter side-band suppression ratio (SBSR) degrades (Fig. S\ref{fig:vernier_design}e). Our requirements for large overcoupling and small length offset result in a Vernier SBSR of a few dB. Measured OPO cold-cavity mode spectra (Fig. \ref{fig:fig1}e.ii) exhibit clear Vernier selectivity, with total Q-factors around $Q_{tot,V}\approx10^6$ (Fig. S\ref{fig:vernier_design}g) and a few dB of sideband suppression. Using a transfer matrix simulation, we infer device parameters. The tuner rings indeed show strong overcoupling of around $10$x ($Q_e\approx95\times10^3$, $Q_i\approx900\times10^3$) near $1600$ nm, resulting in a total round-trip loss of $\approx25$\%. 

\begin{figure*}[t]
  \begin{center}
      \includegraphics[width=\textwidth]{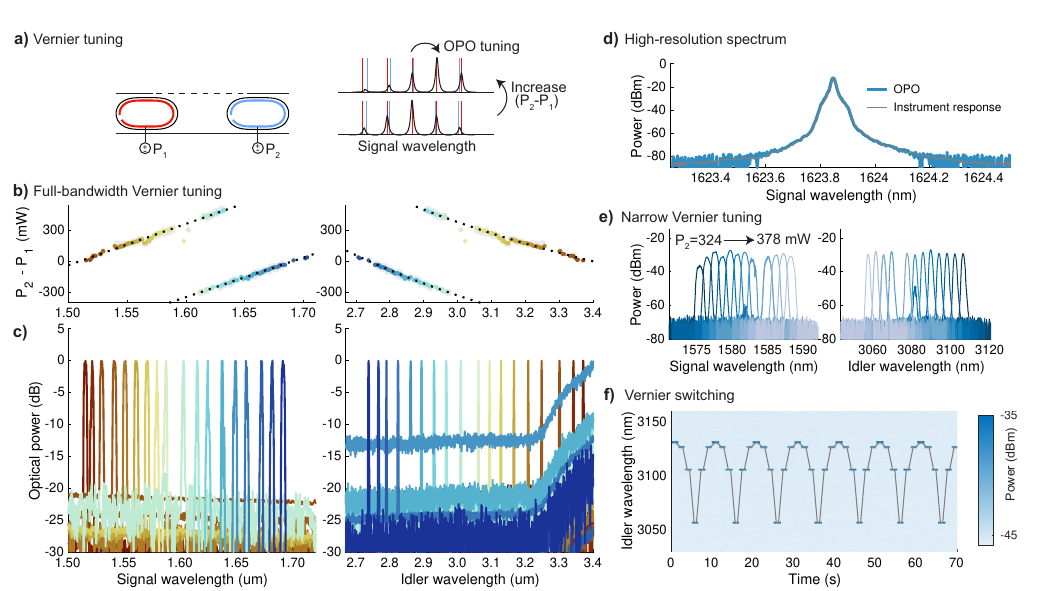}
  \end{center}
 \caption{\textbf{OPO broad wavelength tuning and control.} a) Broad tuning of the Vernier cavity comes from applying differential heater power to the two tuner cavities. In the diagram, red (blue) lines represent mode positions from cavity 1 (2). Black lineshapes indicate the overall transmission of the Vernier filter. (b) Peak OPO signal and idler wavelength tuning for  differential heater power $(P_2-P_1)$. All recorded single-mode OSA peaks shown in gray, while colored points lie closest to fitted linear tuning curves (dotted lines). (c) Selected optical spectra taken at low resolution ($2$~nm). (d) High-resolution ($20$ pm) optical spectrum of the signal wave, which matches the instrument response function taken from a ${<}100$ kHz linewidth tunable diode laser. (e) Optical spectra showing single Vernier mode hop tuning from scanning $P_2$ finely while $P_1$ is held fixed. (f) Optical spectrogram of the idler wavelength tuning controllably and repeatedly over tens of nanometers. The gray line overlay tracks the peak wavelength of each spectrum to guide the eye.}
 \label{fig:fig2}
\end{figure*}

\subsection{Tunability characterization}

Next, we characterize the OPO performance. We fix the wavelength of $1$-\textmu m pump light from a tunable diode laser, then amplify in a fiber amplifier. Because we work with hundreds of milliwatts of on-chip pump power, we operate with quasi-continuous wave pulses to reduce thermal effects and make characterization simpler. The 2-5 \textmu s pulses are generated with an acousto-optic modulator at $10$~kHz repetition rate. The pump light is delivered to the chip's input waveguide with a lensed single-mode fiber ($18.5\pm1$\% edge coupling efficiency). We use either a fluoride multimode fiber to collect signal and idler simultaneously or a lensed single-mode fiber to collect only signal light for highest-resolution spectra. The chip is held at constant temperature during measurements using a thermoelectric cooler. 

Broad tuning is achieved by applying differential heater power $(P_2-P_1)$ to the two tuner cavities (Fig. \ref{fig:fig2}a). Changing $(P_2-P_1)$ by $P_{\mathrm{FSR}}\approx700$~mW---the power needed to shift one cavity by its free spectral range---tunes the output across the full $20$~THz Vernier bandwidth. The individual tuner cavity FSR ($1$~nm at $1.6$~\textmu m signal, $4$~nm at $3$~\textmu m idler) sets the minimum step size of this coarse tuning method.

We test broad tunability by sweeping the on-chip heater powers while holding the pump laser fixed at $1045$~nm. Operating at ${\sim}700$~mW pump power (approximately twice threshold), the OPO produces single-mode emission tunable from $1515$ to $1702$~nm (signal) and $2707$ to $3368$~nm (idler) (Fig. \ref{fig:fig2}b). The output wavelength tunes linearly with $(P_2-P_1)$, spanning two branches separated by the ${\sim}20$~THz Vernier FSR (Fig. \ref{fig:fig2}b). Single-mode spectra are obtained across most of this range (Fig. \ref{fig:fig2}c), though multimode output occasionally appears when the phase heater power $P_{\mathrm{ph}}$ is not optimally set. Small spectral gaps arise from TE/TM mode hybridization \cite{zhangUltrabroadbandIntegratedElectrooptic2025} in the tuner rings (Fig. S\ref{fig:mode_x}), a known effect in x-cut lithium niobate microresonators.

To analyze the output spectra at high resolution, we use a resolution of $20$~pm to measure the emitted signal light coupled into a single mode fiber (Fig. \ref{fig:fig2}d). The OPO signal light looks identical to the instrument response function calibrated using a ${<}100$ kHz linewidth diode laser. Neighboring longitudinal cavity modes, expected to be ${\sim}50$~pm away from the main peak, are not observed, verifying the single-mode emission character. A large OPO side mode suppression ratio (SMSR) of ${>}70$ dB is observed, limited by the spectrum analyzer noise floor. This large SMSR arises despite the passive cavity having only a few dB of mode discrimination. This highlights the more relaxed filter requirements for homogeneously-saturating nonlinear gain compared to semiconductor gain, where SBSRs are limited to ${\sim}40$ dB \cite{agrawalSemiconductorLasers1993}. In parametric gain, pump depletion by the dominant mode uniformly reduces gain for all competing modes, whereas spectral hole burning in semiconductors allows side modes to access unsaturated gain. By energy conservation, the clean optical spectra we observed at signal wavelengths must also be reflected in the mid-infrared idler light, which we cannot measure directly at high resolution because of multimode fiber collection.

Next we verify narrow wavelength tuning over single Vernier mode hops. Here, $P_2$ is swept from $324$ to $378$ mW while holding $P_1$ fixed. We plot a typical narrow range of tuning in Fig. \ref{fig:fig2}e. Though small spectral gaps can sometimes be observed, the OPO most often tunes cleanly over single Vernier mode hops (${\sim}125$ GHz) as expected. 

Concluding the broad tunability study, we test the device's repeatability and reconfigurability. To do so, $P_1$ is held fixed while $P_2$ programmably switches between four different settings to verify mid-infrared switching performance with both few-nm and tens-of-nm jumps. We sweep our optical spectrum as fast as possible ($0.5$s per scan) over a $200$ nm range. After accounting for a ${\sim}0.5$s thermal settling time required after the OPO programming switches, we plot the optical spectrogram in Fig. \ref{fig:fig2}f. As seen, the mid-infrared wide tuning is highly repeatable and switchable over second timescales, always landing within the same Vernier mode hop window. These switching times, which we attribute to the slow seconds-long impulse response of our temperature control system, have been brought down to sub-ms levels in optimized designs on TFLN \cite{liuHighlyEfficientThermooptic2020, qiLowlossHighlyTunable2025}. These switching measurements already highlight our Vernier OPO architecture's ability to support simple yet robust wavelength reconfiguration.

\begin{figure*}[t]
  \begin{center}
      \includegraphics[width=\textwidth]{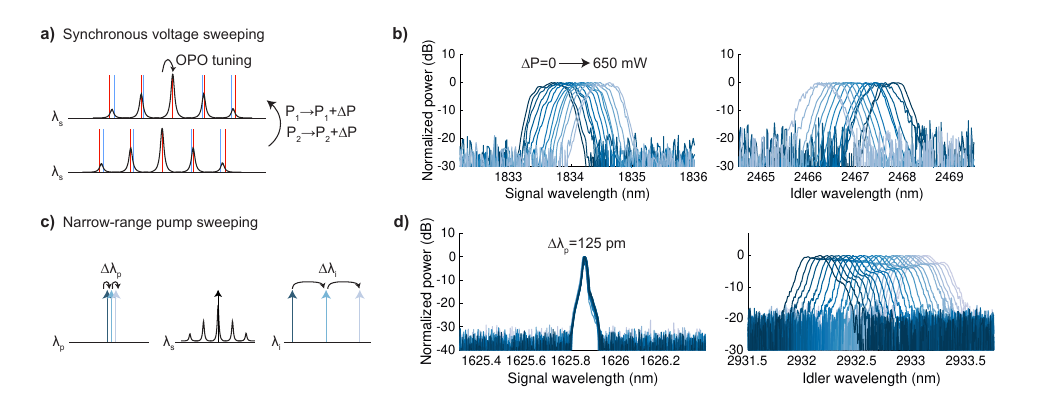}
  \end{center}
 \caption{\textbf{Fine wavelength control.} a) Tuning the heater cavities simultaneously shifts the Vernier mode with sub-nm control. b) Optical spectra of signal and idler during synchronous addition of heater power $\Delta P$ to both tuner cavities. c) Mode-hop-free mid-infrared tuning can be achieved by tuning the pump narrowly while the signal remains pinned at a Vernier mode. d) Optical spectra of signal and idler taken while piezo-tuning the pump over $125$ pm ($34$ GHz).  The signal (taken with $20$-pm resolution from a single-mode fiber) stays pinned at single Vernier mode while the idler (taken with a $100$-pm resolution from a multi-mode fiber) tunes continuously.}
 \label{fig:fig3}
\end{figure*}

Because accessing specific mid-infrared spectral features requires high-resolution wavelength control, we also test the device's ability to tune in between Vernier mode hop windows. First, we use the most common method to narrowly tune a Vernier laser: tuning the heaters on both tuner cavities synchronously, which drags the Vernier filter peak without a Vernier mode hop (Fig. \ref{fig:fig3}a). Tuning across the entire intra-mode-hop region requires applying $P_{\mathrm{FSR}}$ to both tuner cavities. We verify Vernier fine tuning ability in Fig. \ref{fig:fig3}b. For this demonstration, we use a second device on the chip that oscillates at signal wavelengths from $1.7$-$1.9$~\textmu m. This device exhibited tuning behavior consistent with the primary device before the latter was damaged. By simultaneously adding $\Delta P = 650$~mW of heater power to each tuner cavity, we successfully observe ${\sim}1$~nm of finely-resolved signal redshift, corresponding to ${\sim}2$~nm of idler blueshift. True mode hop free tuning can be obtained by adding $P_{\mathrm{FSR}}$ to the phase section as well. This is not possible in our current device, because the long phase section's large resistance requires large voltage ($120$ V) to generate $P_{\mathrm{FSR}}$.

Our OPO architecture also allows us to demonstrate fine mode hop free tuning in a simple and practical way by narrowly tuning the pump laser. This method exploits the energy conservation between pump, signal, and idler and the degrees of freedom provided by the chosen singly-resonant architecture, which constrains the wavelength of signal, but not pump and idler (Fig. \ref{fig:fig3}c). While the Vernier cavity defines the signal wavelength exactly, frequency translation of the pump freely transfers to the idler. This directly utilizes the ability of nearly all compact semiconductor lasers to tune mode-hop-free in narrow ranges. Instead of requiring three synchronously-controlled voltage signals as in the synchronous Vernier tuning method above, here we only require one signal on the pump laser. We piezo-tune our pump laser approximately over its maximum mode-hop-free tuning range ($35$~GHz, or $125$~pm at $1045$~nm) (Fig. \ref{fig:fig3}d). At each fine step of the pump laser wavelength, the signal remains locked in the same longitudinal mode as expected. Meanwhile, the idler scans finely over a nanometer at $2.9$~\textmu m, directly following the pump's frequency translation.

\begin{figure}[t]
  \begin{center}
      \includegraphics[width=0.5\textwidth]{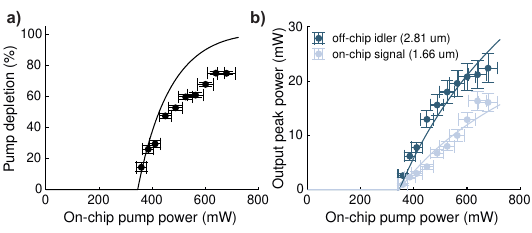}
  \end{center}
 \caption{\textbf{Power characterization.} We sweep the on-chip peak power of a rectangular pump pulse ($1045$ nm, $10$ kHz repetition rate, $5$\% duty cycle). For each power step, we measure maximum values during pump turn-on of: a) Pump depletion and b) output on-chip signal and off-chip idler powers. Solid line: expected trends from modeling an ideal low loss singly-resonant OPO.}
 \label{fig:fig4}
\end{figure}

\subsection{Power output characterization}

Finally, we characterize device threshold and output power by sweeping the on-chip pump pulse power (Fig. \ref{fig:fig4}). We measure the pump's depletion (Fig. \ref{fig:fig4}a) by tuning the device's heaters to either oscillate at $1.66$~\textmu m/$2.81$~\textmu m signal/idler wavelengths or not oscillate at all, which provides the control signal that we normalize to. The oscillation on/off states have slightly different pump throughputs, which we normalize out but leads to some noise in the extracted pump depletion. We also report the output signal and idler power in Fig. \ref{fig:fig4}b. We report off-chip idler power, which is calibrated by measuring off-chip mid-infrared light produced by difference frequency generation using a thermal power sensor. With more characterization capability at signal wavelengths, we can be more specific and report on-chip power. For both pump depletion and output power data, we report maximal values over the duration of the $5$~\textmu s pump pulse. The device begins oscillating at $P_{th}\approx380$~mW. As pump power increases above threshold, the pump depletion and output powers roughly follow the expected dependences of an ideal singly-resonant OPO. At the maximum on-chip pump power we send (${\sim}$700~mW), the pump depletion approaches $75$\% while the output idler power reaches $22$~mW off-chip. The signal has smaller measured power, as expected, due to its weaker cavity outcoupling. The maximal pump-to-idler conversion efficiency is ${\sim}3$\%, around $8$x lower than expected based on the maximum measured pump depletion and quantum defect-limited conversion efficiency (${\sim}33$\%). We attribute most of these efficiency losses to the idler extraction coupler, which we found from simulation to have suboptimal extraction efficiency. 

\section{Discussion}

\begin{figure}[t]
  \begin{center}
      \includegraphics[width=0.5\textwidth]{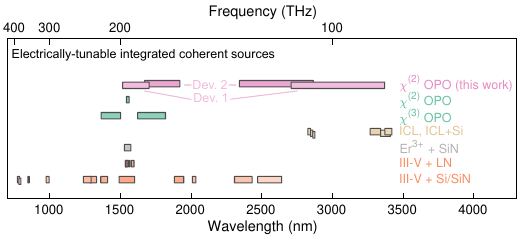}
  \end{center}
 \caption{\textbf{Overview of electrically-tunable integrated coherent sources.} We compare the electrical tuning range of two separate devices from this work on the same chip, each operated with a fixed-wavelength pump at constant temperature, to other recent demonstrations. We group previous demonstrations by gain platform: III-V integrated with silicon \cite{mortonIntegratedCoherentTunable2022, siaAnalysisCompactSilicon2020, guoEbandWidelyTunable2023, malikWidelyTunableHeterogeneously2020, komljenovicWidelyTunableNarrowLinewidth2015, xieHighefficiencyTunableLasers2025, briansiaSubkHzLinewidthHybrid2020, weiUltrabroadbandTunableGaSbsilicon2025, weiWidelyTunableShortwave2025} or silicon nitride \cite{tranExtendingSpectrumFully2022a, castroIntegratedModehopfreeTunable2025, nejadriahiSub100HzIntrinsic2024, ojanenWidelyTunable2472642023} photonics; III-V integrated with lithium niobate \cite{liIntegratedPockelsLaser2022a, opdebeeckIIIVonlithiumNiobate2021a, lufungulaTunableDualWavelength2023}; Er$^{3+}$ ions integrated with silicon nitride \cite{liMonolithicallyIntegratedErbiumdoped2018}; interband cascade laser gain with tunability from directly-etched V-coupled cavities at a single temperature \cite{ yangMidinfraredWidelyTunable2020, wangWidelyTunableSinglemode2024} or silicon photonics \cite{shimTunableSinglemodeChipscale2021}; $\chi^{(3)}$ OPO at a single temperature with $20$ GHz pump wavelength tuning \cite{pidgaykoVoltagetunableOpticalParametric2023}; and $\chi^{(2)}$ OPOs at a single temperature/pump wavelength \cite{kellnerLowThresholdIntegrated2025}. Different references under the same category are shaded differently, while different device realizations under the same reference are vertically offset slightly.}
\label{fig:fig5}
\end{figure}

In summary, we demonstrate a conceptually simple yet powerful integrated OPO architecture, allowing a single device pumped with a fixed-wavelength near-infrared laser to generate $22$~THz of electrically-tunable mid-infrared near $3$~\textmu m. The output mid-infrared can be controlled across three spectral scales: coarse tuning across the full $22$~THz gain bandwidth via differential heater power, discrete $125$~GHz steps between Vernier modes, and continuous sub-$125$~GHz tuning within each Vernier window. The mid-infrared tuning is reliable and repeatable, while retaining excellent spectral purity (SMSR ${>}70$~dB), with tens of milliwatts of off-chip power. These experiments verify the potential of integrated nonlinear devices to create powerful broadly tunable mid-infrared sources.

Reducing the optical threshold will be important to enable pump laser integration with the OPO. Several straightforward improvements can bring the current $380$~mW threshold to the tens of milliwatt level: increasing the tuner cavity overcoupling ratio from 10x to 20x reduces round-trip loss by 2x (already achieved on a separate fabrication run); adaptive poling techniques \cite{chenAdaptedPolingBreak2024} can recover the 1.7x gain reduction from film thickness variations (see Methods); and the largest reductions can come from resonating the pump in an additional cavity \cite{deanLowpowerIntegratedOptical2026}, which is feasible with a modification of our architecture. Meanwhile, the electrical power required to tune over the whole bandwidth ($P_{\mathrm{FSR}}\approx700$~mW) can be reduced by 25x by utilizing previously-demonstrated thermal isolation trenching strategies \cite{liuHighlyEfficientThermooptic2020, qiLowlossHighlyTunable2025}. 

To put these results in context, we compare the tuning ranges of two devices from this work to those of other electrically-tunable integrated coherent sources (Fig. \ref{fig:fig5}). We include the device primarily focused on in the paper (dev. 1, $2.7$-$3.4$~\textmu m idler) and a second device on the same chip (dev. 2, $2.35$-$2.9$~\textmu m idler). Combined, the two devices' electrical tuning at a single pump wavelength spans $1.5$-$3.4$~\textmu m with a $400$~nm gap at degeneracy. While broadly tunable (${>}10$~THz) lasers are now widely demonstrated from the visible to near-infrared using III/V + silicon or silicon nitride photonics, spectral coverage beyond the telecom L-band (${>}1.65$~\textmu m) becomes much sparser. Especially above $2$~\textmu m, the best option becomes interband cascade laser gain, which has been incorporated a few times into ${\sim}100$~nm tunable devices. Fig. \ref{fig:fig5} clearly highlights the advantage of using a $\chi^{(2)}$ OPO to reach the difficult spectral ranges from $2$-$3.5$~\textmu m.

\bibliography{refs}

@book{agrawalSemiconductorLasers1993,
  title = {Semiconductor {{Lasers}}},
  author = {Agrawal, Govind P. and Dutta, Niloy K.},
  year = 1993,
  edition = {Second Edition},
  publisher = {Springer US},
  address = {Boston, MA},
  doi = {10.1007/978-1-4613-0481-4},
  isbn = {978-1-4613-0481-4},
  langid = {english}
}

@article{boesLithiumNiobatePhotonics2023,
  title = {Lithium Niobate Photonics: {{Unlocking}} the Electromagnetic Spectrum},
  shorttitle = {Lithium Niobate Photonics},
  author = {Boes, Andreas and Chang, Lin and Langrock, Carsten and Yu, Mengjie and Zhang, Mian and Lin, Qiang and Lon{\v c}ar, Marko and Fejer, Martin and Bowers, John and Mitchell, Arnan},
  year = 2023,
  month = jan,
  journal = {Science},
  volume = {379},
  number = {6627},
  pages = {eabj4396},
  issn = {0036-8075, 1095-9203},
  doi = {10.1126/science.abj4396},
  urldate = {2025-09-26},
  langid = {english}
}

@article{briansiaSubkHzLinewidthHybrid2020,
  title = {Sub-{{kHz}} Linewidth, Hybrid {{III-V}}/Silicon Wavelength-Tunable Laser Diode Operating at the Application-Rich 1647-1690 Nm},
  author = {Brian Sia, Jia Xu and Li, Xiang and Wang, Wanjun and Qiao, Zhongliang and Guo, Xin and Zhou, Jin and Littlejohns, Callum G. and Liu, Chongyang and Reed, Graham T. and Wang, Hong},
  year = 2020,
  month = aug,
  journal = {Optics Express},
  volume = {28},
  number = {17},
  pages = {25215},
  issn = {1094-4087},
  doi = {10.1364/OE.400666},
  urldate = {2025-08-17},
  langid = {english}
}

@article{castroIntegratedModehopfreeTunable2025,
  title = {Integrated Mode-Hop-Free Tunable Lasers at 780 Nm for Chip-Scale Classical and Quantum Photonic Applications},
  author = {Castro, Joshua E. and {Nolasco-Martinez}, Eber and Pintus, Paolo and Zhang, Zeyu and Shen, Boqiang and Morin, Theodore and Thiel, Lillian and Steiner, Trevor J. and Lewis, Nicholas and Patel, Sahil D. and Bowers, John E. and Weld, David M. and Moody, Galan},
  year = 2025,
  month = mar,
  journal = {APL Photonics},
  volume = {10},
  number = {3},
  pages = {036102},
  issn = {2378-0967},
  doi = {10.1063/5.0232377},
  urldate = {2025-08-17},
  langid = {english}
}

@article{chenAdaptedPolingBreak2024,
  title = {Adapted Poling to Break the Nonlinear Efficiency Limit in Nanophotonic Lithium Niobate Waveguides},
  author = {Chen, Pao-Kang and Briggs, Ian and Cui, Chaohan and Zhang, Liang and Shah, Manav and Fan, Linran},
  year = 2024,
  month = jan,
  journal = {Nature Nanotechnology},
  volume = {19},
  number = {1},
  pages = {44--50},
  issn = {1748-3387, 1748-3395},
  doi = {10.1038/s41565-023-01525-w},
  urldate = {2025-09-23},
  langid = {english}
}

@article{chikkaraddySinglemoleculeMidinfraredSpectroscopy2023,
  title = {Single-Molecule Mid-Infrared Spectroscopy and Detection through Vibrationally Assisted Luminescence},
  author = {Chikkaraddy, Rohit and Arul, Rakesh and Jakob, Lukas A. and Baumberg, Jeremy J.},
  year = 2023,
  month = oct,
  journal = {Nature Photonics},
  volume = {17},
  number = {10},
  pages = {865--871},
  issn = {1749-4885, 1749-4893},
  doi = {10.1038/s41566-023-01263-4},
  urldate = {2025-09-25},
  langid = {english}
}

@article{chilesMidinfraredIntegratedWaveguide2014,
  title = {Mid-Infrared Integrated Waveguide Modulators Based on Silicon-on-Lithium-Niobate Photonics},
  author = {Chiles, Jeff and Fathpour, Sasan},
  year = 2014,
  month = nov,
  journal = {Optica},
  volume = {1},
  number = {5},
  pages = {350},
  issn = {2334-2536},
  doi = {10.1364/OPTICA.1.000350},
  urldate = {2026-02-09},
  copyright = {https://doi.org/10.1364/OA\_License\_v1\#VOR-OA},
  langid = {english}
}

@article{deanLowpowerIntegratedOptical2026,
  title = {Low-Power Integrated Optical Amplification through Second-Harmonic Resonance},
  author = {Dean, Devin J. and Park, Taewon and Stokowski, Hubert S. and Qi, Luke and Robison, Sam and Hwang, Alexander Y. and Herrmann, Jason F. and Fejer, Martin M. and {Safavi-Naeini}, Amir H.},
  year = 2026,
  month = jan,
  journal = {Nature},
  volume = {649},
  number = {8099},
  pages = {1159--1164},
  issn = {0028-0836, 1476-4687},
  doi = {10.1038/s41586-025-09959-z},
  urldate = {2026-02-09},
  langid = {english}
}

@article{duttNonlinearQuantumPhotonics2024,
  title = {Nonlinear and Quantum Photonics Using Integrated Optical Materials},
  author = {Dutt, Avik and Mohanty, Aseema and Gaeta, Alexander L. and Lipson, Michal},
  year = 2024,
  month = may,
  journal = {Nature Reviews Materials},
  volume = {9},
  number = {5},
  pages = {321--346},
  issn = {2058-8437},
  doi = {10.1038/s41578-024-00668-z},
  urldate = {2025-09-26},
  langid = {english}
}

@article{fengIntegratedLithiumNiobate2024,
  title = {Integrated Lithium Niobate Microwave Photonic Processing Engine},
  author = {Feng, Hanke and Ge, Tong and Guo, Xiaoqing and Wang, Benshan and Zhang, Yiwen and Chen, Zhaoxi and Zhu, Sha and Zhang, Ke and Sun, Wenzhao and Huang, Chaoran and Yuan, Yixuan and Wang, Cheng},
  year = 2024,
  month = mar,
  journal = {Nature},
  volume = {627},
  number = {8002},
  pages = {80--87},
  issn = {0028-0836, 1476-4687},
  doi = {10.1038/s41586-024-07078-9},
  urldate = {2025-09-26},
  langid = {english}
}

@article{fuchsbergerContinuouslyWidelyTunable2025,
  title = {Continuously and Widely Tunable Semiconductor Ring Lasers},
  author = {Fuchsberger, Johannes and Letsou, Theodore P. and Kazakov, Dmitry and Szedlak, Rolf and Capasso, Federico and Schwarz, Benedikt},
  year = 2025,
  month = jul,
  journal = {Optica},
  volume = {12},
  number = {7},
  pages = {985},
  issn = {2334-2536},
  doi = {10.1364/OPTICA.559884},
  urldate = {2025-08-17},
  langid = {english}
}

@article{guoEbandWidelyTunable2023,
  title = {E-Band Widely Tunable, Narrow Linewidth Heterogeneous Laser on Silicon},
  author = {Guo, Joel and Xiang, Chao and Morin, Theodore J. and Peters, Jonathan D. and Chang, Lin and Bowers, John E.},
  year = 2023,
  month = apr,
  journal = {APL Photonics},
  volume = {8},
  number = {4},
  pages = {046114},
  issn = {2378-0967},
  doi = {10.1063/5.0133040},
  urldate = {2025-08-17},
  langid = {english}
}

@article{heimHybridIntegratedUltralow2025,
  title = {Hybrid Integrated Ultra-Low Linewidth Coil Stabilized Isolator-Free Widely Tunable External Cavity Laser},
  author = {Heim, David A. S. and Bose, Debapam and Liu, Kaikai and Isichenko, Andrei and Blumenthal, Daniel J.},
  year = 2025,
  month = jul,
  journal = {Nature Communications},
  volume = {16},
  number = {1},
  pages = {5944},
  issn = {2041-1723},
  doi = {10.1038/s41467-025-61122-4},
  urldate = {2025-09-26},
  langid = {english}
}

@article{huHighefficiencyBroadbandOnchip2022a,
  title = {High-Efficiency and Broadband on-Chip Electro-Optic Frequency Comb Generators},
  author = {Hu, Yaowen and Yu, Mengjie and Buscaino, Brandon and Sinclair, Neil and Zhu, Di and Cheng, Rebecca and {Shams-Ansari}, Amirhassan and Shao, Linbo and Zhang, Mian and Kahn, Joseph M. and Lon{\v c}ar, Marko},
  year = 2022,
  month = oct,
  journal = {Nature Photonics},
  volume = {16},
  number = {10},
  pages = {679--685},
  issn = {1749-4885, 1749-4893},
  doi = {10.1038/s41566-022-01059-y},
  urldate = {2026-02-09},
  langid = {english}
}

@article{hwangMidinfraredSpectroscopyBroadly2023a,
  title = {Mid-Infrared Spectroscopy with a Broadly Tunable Thin-Film Lithium Niobate Optical Parametric Oscillator},
  author = {Hwang, Alexander Y. and Stokowski, Hubert S. and Park, Taewon and Jankowski, Marc and McKenna, Timothy P. and Langrock, Carsten and Mishra, Jatadhari and Ansari, Vahid and Fejer, Martin M. and {Safavi-Naeini}, Amir H.},
  year = 2023,
  month = nov,
  journal = {Optica},
  volume = {10},
  number = {11},
  pages = {1535},
  issn = {2334-2536},
  doi = {10.1364/OPTICA.502487},
  urldate = {2025-09-26},
  langid = {english}
}

@article{jankowskiDispersionengineeredH2Nanophotonics2021b,
  title = {Dispersion-Engineered {$\chi$}(2) Nanophotonics: A Flexible Tool for Nonclassical Light},
  shorttitle = {Dispersion-Engineered {$\chi$}(2) Nanophotonics},
  author = {Jankowski, Marc and Mishra, Jatadhari and Fejer, M M},
  year = 2021,
  month = oct,
  journal = {Journal of Physics: Photonics},
  volume = {3},
  number = {4},
  pages = {042005},
  issn = {2515-7647},
  doi = {10.1088/2515-7647/ac1729},
  urldate = {2025-09-26}
}

@article{jiMultimodalityIntegratedMicroresonators2024,
  title = {Multimodality Integrated Microresonators Using the {{Moir\'e}} Speedup Effect},
  author = {Ji, Qing-Xin and Liu, Peng and Jin, Warren and Guo, Joel and Wu, Lue and Yuan, Zhiquan and Peters, Jonathan and Feshali, Avi and Paniccia, Mario and Bowers, John E. and Vahala, Kerry J.},
  year = 2024,
  month = mar,
  journal = {Science},
  volume = {383},
  number = {6687},
  pages = {1080--1083},
  issn = {0036-8075, 1095-9203},
  doi = {10.1126/science.adk9429},
  urldate = {2026-02-09},
  langid = {english}
}

@article{kazakovDrivenBrightSolitons2025,
  title = {Driven Bright Solitons on a Mid-Infrared Laser Chip},
  author = {Kazakov, Dmitry and Letsou, Theodore P. and Piccardo, Marco and Columbo, Lorenzo L. and Brambilla, Massimo and Prati, Franco and Dal Cin, Sandro and Beiser, Maximilian and Opa{\v c}ak, Nikola and Ratra, Pawan and Pushkarsky, Michael and Caffey, David and Day, Timothy and Lugiato, Luigi A. and Schwarz, Benedikt and Capasso, Federico},
  year = 2025,
  month = may,
  journal = {Nature},
  volume = {641},
  number = {8061},
  pages = {83--89},
  issn = {0028-0836, 1476-4687},
  doi = {10.1038/s41586-025-08853-y},
  urldate = {2025-09-25},
  langid = {english}
}

@article{kellnerLowThresholdIntegrated2025,
  title = {Low Threshold Integrated Optical Parametric Oscillator with a Compact {{Bragg}} Resonator},
  author = {Kellner, Jost and Sabatti, Alessandra and Maeder, Andreas and Grange, Rachel},
  year = 2025,
  month = may,
  journal = {Optica},
  volume = {12},
  number = {5},
  pages = {702},
  issn = {2334-2536},
  doi = {10.1364/OPTICA.558804},
  urldate = {2025-08-17},
  langid = {english}
}

@article{komljenovicWidelyTunableNarrowLinewidth2015,
  title = {Widely {{Tunable Narrow-Linewidth Monolithically Integrated External-Cavity Semiconductor Lasers}}},
  author = {Komljenovic, Tin and Srinivasan, Sudharsanan and Norberg, Erik and Davenport, Michael and Fish, Gregory and Bowers, John E.},
  year = 2015,
  month = nov,
  journal = {IEEE Journal of Selected Topics in Quantum Electronics},
  volume = {21},
  number = {6},
  pages = {214--222},
  issn = {1077-260X, 1558-4542},
  doi = {10.1109/JSTQE.2015.2422752},
  urldate = {2025-08-17},
  copyright = {https://ieeexplore.ieee.org/Xplorehelp/downloads/license-information/IEEE.html}
}

@article{ledezmaOctavespanningTunableInfrared2023a,
  title = {Octave-Spanning Tunable Infrared Parametric Oscillators in Nanophotonics},
  author = {Ledezma, Luis and Roy, Arkadev and Costa, Luis and Sekine, Ryoto and Gray, Robert and Guo, Qiushi and Nehra, Rajveer and Briggs, Ryan M. and Marandi, Alireza},
  year = 2023,
  month = jul,
  journal = {Science Advances},
  volume = {9},
  number = {30},
  pages = {eadf9711},
  issn = {2375-2548},
  doi = {10.1126/sciadv.adf9711},
  urldate = {2025-09-26},
  langid = {english}
}

@article{liangModulatedRingdownComb2025,
  title = {Modulated Ringdown Comb Interferometry for Sensing of Highly Complex Gases},
  author = {Liang, Qizhong and Bisht, Apoorva and Scheck, Andrew and Schunemann, Peter G. and Ye, Jun},
  year = 2025,
  month = feb,
  journal = {Nature},
  volume = {638},
  number = {8052},
  pages = {941--948},
  issn = {0028-0836, 1476-4687},
  doi = {10.1038/s41586-024-08534-2},
  urldate = {2025-09-25},
  langid = {english}
}

@article{liIntegratedPockelsLaser2022a,
  title = {Integrated {{Pockels}} Laser},
  author = {Li, Mingxiao and Chang, Lin and Wu, Lue and Staffa, Jeremy and Ling, Jingwei and Javid, Usman A. and Xue, Shixin and He, Yang and {Lopez-rios}, Raymond and Morin, Theodore J. and Wang, Heming and Shen, Boqiang and Zeng, Siwei and Zhu, Lin and Vahala, Kerry J. and Bowers, John E. and Lin, Qiang},
  year = 2022,
  month = sep,
  journal = {Nature Communications},
  volume = {13},
  number = {1},
  pages = {5344},
  issn = {2041-1723},
  doi = {10.1038/s41467-022-33101-6},
  urldate = {2025-08-17},
  langid = {english}
}

@article{liMonolithicallyIntegratedErbiumdoped2018,
  title = {Monolithically Integrated Erbium-Doped Tunable Laser on a {{CMOS-compatible}} Silicon Photonics Platform},
  author = {Li, Nanxi and Vermeulen, Diedrik and Su, Zhan and Magden, Emir Salih and Xin, Ming and Singh, Neetesh and Ruocco, Alfonso and Notaros, Jelena and Poulton, Christopher V. and Timurdogan, Erman and Baiocco, Christopher and Watts, Michael R.},
  year = 2018,
  month = jun,
  journal = {Optics Express},
  volume = {26},
  number = {13},
  pages = {16200},
  issn = {1094-4087},
  doi = {10.1364/OE.26.016200},
  urldate = {2025-08-17},
  langid = {english}
}

@article{liuHighlyEfficientThermooptic2020,
  title = {Highly Efficient Thermo-Optic Tunable Micro-Ring Resonator Based on an {{LNOI}} Platform},
  author = {Liu, Xiaoyue and Ying, Pan and Zhong, Xuming and Xu, Jian and Han, Ya and Yu, Siyuan and Cai, Xinlun},
  year = 2020,
  month = nov,
  journal = {Optics Letters},
  volume = {45},
  number = {22},
  pages = {6318},
  issn = {0146-9592, 1539-4794},
  doi = {10.1364/OL.410192},
  urldate = {2025-09-26},
  langid = {english}
}

@article{longNanosecondTimeresolvedDualcomb2024,
  title = {Nanosecond Time-Resolved Dual-Comb Absorption Spectroscopy},
  author = {Long, David A. and Cich, Matthew J. and Mathurin, Carl and Heiniger, Adam T. and Mathews, Garrett C. and Frymire, Augustine and Rieker, Gregory B.},
  year = 2024,
  month = feb,
  journal = {Nature Photonics},
  volume = {18},
  number = {2},
  pages = {127--131},
  issn = {1749-4885, 1749-4893},
  doi = {10.1038/s41566-023-01316-8},
  urldate = {2026-02-09},
  langid = {english}
}

@article{luEmergingIntegratedLaser2024,
  title = {Emerging Integrated Laser Technologies in the Visible and Short Near-Infrared Regimes},
  author = {Lu, Xiyuan and Chang, Lin and Tran, Minh A. and Komljenovic, Tin and Bowers, John E. and Srinivasan, Kartik},
  year = 2024,
  month = oct,
  journal = {Nature Photonics},
  volume = {18},
  number = {10},
  pages = {1010--1023},
  issn = {1749-4885, 1749-4893},
  doi = {10.1038/s41566-024-01529-5},
  urldate = {2025-09-26},
  langid = {english}
}

@inproceedings{lufungulaTunableDualWavelength2023,
  title = {Tunable Dual Wavelength Laser on Thin Film Lithium Niobate},
  booktitle = {2023 {{IEEE Photonics Conference}} ({{IPC}})},
  author = {Lufungula, Isaac Luntadila and Mayor, Felix M. and Herrmann, Jason F. and Park, Taewon and Stokowski, Hubert S. and Hwang, Alexander Y. and De Beeck, Camiel Op and Atalar, Okan and Jiang, Wentao and Kuyken, Bart and {Safavi-Naeini}, Amir H.},
  year = 2023,
  month = nov,
  pages = {1--2},
  publisher = {IEEE},
  address = {Orlando, FL, USA},
  doi = {10.1109/IPC57732.2023.10360651},
  urldate = {2025-08-17},
  copyright = {https://doi.org/10.15223/policy-029},
  isbn = {979-8-3503-4722-7}
}

@article{luPhotonicIntegratedCircuit2026,
  title = {Photonic Integrated Circuit Optical Parametric Oscillators},
  author = {Lu, Xiyuan and Gray, Robert M. and Stone, Jordan and Zhou, Selina and Englebert, Nicolas and Marandi, Alireza and Srinivasan, Kartik},
  year = 2026,
  month = jan,
  journal = {Optica},
  volume = {13},
  number = {1},
  pages = {11},
  issn = {2334-2536},
  doi = {10.1364/OPTICA.572694},
  urldate = {2026-01-29},
  langid = {english}
}

@article{malikWidelyTunableHeterogeneously2020,
  title = {Widely Tunable, Heterogeneously Integrated Quantum-Dot {{O-band}} Lasers on Silicon},
  author = {Malik, Aditya and Guo, Joel and Tran, Minh A. and Kurczveil, Geza and Liang, Di and Bowers, John E.},
  year = 2020,
  month = oct,
  journal = {Photonics Research},
  volume = {8},
  number = {10},
  pages = {1551},
  issn = {2327-9125},
  doi = {10.1364/PRJ.394726},
  urldate = {2025-08-17},
  langid = {english}
}

@article{mishraUltrabroadbandMidinfraredGeneration2022b,
  title = {Ultra-Broadband Mid-Infrared Generation in Dispersion-Engineered Thin-Film Lithium Niobate},
  author = {Mishra, Jatadhari and Jankowski, Marc and Hwang, Alexander Y. and Stokowski, Hubert S. and McKenna, Timothy P. and Langrock, Carsten and Ng, Edwin and Heydari, David and Mabuchi, Hideo and {Safavi-Naeini}, Amir H. and Fejer, M. M.},
  year = 2022,
  month = aug,
  journal = {Optics Express},
  volume = {30},
  number = {18},
  pages = {32752},
  issn = {1094-4087},
  doi = {10.1364/OE.467580},
  urldate = {2025-09-26},
  langid = {english}
}

@article{montesUltracoherentSignalOutput2004,
  title = {Ultra-Coherent Signal Output from an Incoherent Cw-Pumped Singly Resonant Optical Parametric Oscillator},
  author = {Montes, Carlos and Picozzi, Antonio and Gallo, Katia},
  year = 2004,
  month = jul,
  journal = {Optics Communications},
  volume = {237},
  number = {4-6},
  pages = {437--449},
  issn = {00304018},
  doi = {10.1016/j.optcom.2004.04.017},
  urldate = {2026-02-09},
  copyright = {https://www.elsevier.com/tdm/userlicense/1.0/},
  langid = {english}
}

@article{mortonIntegratedCoherentTunable2022,
  title = {Integrated {{Coherent Tunable Laser}} ({{ICTL}}) {{With Ultra-Wideband Wavelength Tuning}} and {{Sub-100 Hz Lorentzian Linewidth}}},
  author = {Morton, Paul A. and Xiang, Chao and Khurgin, Jacob B. and Morton, Christopher D. and Tran, Minh and Peters, Jon and Guo, Joel and Morton, Michael J. and Bowers, John E.},
  year = 2022,
  month = mar,
  journal = {Journal of Lightwave Technology},
  volume = {40},
  number = {6},
  pages = {1802--1809},
  issn = {0733-8724, 1558-2213},
  doi = {10.1109/JLT.2021.3127155},
  urldate = {2025-08-17},
  copyright = {https://creativecommons.org/licenses/by/4.0/legalcode}
}

@article{nejadriahiSub100HzIntrinsic2024,
  title = {Sub-100 {{Hz}} Intrinsic Linewidth 852 Nm Silicon Nitride External Cavity Laser},
  author = {Nejadriahi, Hani and Kittlaus, Eric and Bose, Debapam and Chauhan, Nitesh and Wang, Jiawei and Fradet, Mathieu and Bagheri, Mahmood and Isichenko, Andrei and Heim, David and Forouhar, Siamak and Blumenthal, Daniel J.},
  year = 2024,
  month = dec,
  journal = {Optics Letters},
  volume = {49},
  number = {24},
  pages = {7254},
  issn = {0146-9592, 1539-4794},
  doi = {10.1364/OL.543307},
  urldate = {2025-08-17},
  langid = {english}
}

@article{nitissOpticallyReconfigurableQuasiphasematching2022,
  title = {Optically Reconfigurable Quasi-Phase-Matching in Silicon Nitride Microresonators},
  author = {Nitiss, Edgars and Hu, Jianqi and Stroganov, Anton and Br{\`e}s, Camille-Sophie},
  year = 2022,
  month = feb,
  journal = {Nature Photonics},
  volume = {16},
  number = {2},
  pages = {134--141},
  issn = {1749-4885, 1749-4893},
  doi = {10.1038/s41566-021-00925-5},
  urldate = {2026-02-09},
  langid = {english}
}

@article{ojanenWidelyTunable2472642023,
  title = {Widely {{Tunable}} (2.47--2.64 \textmu m) {{Hybrid Laser Based}} on {{GaSb}}/{{GaInAsSb Quantum}}-{{Wells}} and a {{Low}}-{{Loss Si}}{\textsubscript{3}} {{N}}{\textsubscript{4}} {{Photonic Integrated Circuit}}},
  author = {Ojanen, Samu-Pekka and Viheri{\"a}l{\"a}, Jukka and Zia, Nouman and Koivusalo, Eero and Hilska, Joonas and Tuorila, Heidi and Guina, Mircea},
  year = 2023,
  month = jul,
  journal = {Laser \& Photonics Reviews},
  volume = {17},
  number = {7},
  pages = {2201028},
  issn = {1863-8880, 1863-8899},
  doi = {10.1002/lpor.202201028},
  urldate = {2025-10-16},
  langid = {english}
}

@article{opdebeeckIIIVonlithiumNiobate2021a,
  title = {{{III}}/{{V-on-lithium}} Niobate Amplifiers and Lasers},
  author = {Op De Beeck, Camiel and Mayor, Felix M. and Cuyvers, Stijn and Poelman, Stijn and Herrmann, Jason F. and Atalar, Okan and McKenna, Timothy P. and Haq, Bahawal and Jiang, Wentao and Witmer, Jeremy D. and Roelkens, Gunther and {Safavi-Naeini}, Amir H. and Van Laer, Rapha{\"e}l and Kuyken, Bart},
  year = 2021,
  month = oct,
  journal = {Optica},
  volume = {8},
  number = {10},
  pages = {1288},
  issn = {2334-2536},
  doi = {10.1364/OPTICA.438620},
  urldate = {2025-08-17},
  langid = {english}
}

@article{perezHighperformanceKerrMicroresonator2023,
  title = {High-Performance {{Kerr}} Microresonator Optical Parametric Oscillator on a Silicon Chip},
  author = {Perez, Edgar F. and Moille, Gr{\'e}gory and Lu, Xiyuan and Stone, Jordan and Zhou, Feng and Srinivasan, Kartik},
  year = 2023,
  month = jan,
  journal = {Nature Communications},
  volume = {14},
  number = {1},
  pages = {242},
  issn = {2041-1723},
  doi = {10.1038/s41467-022-35746-9},
  urldate = {2025-09-26},
  langid = {english}
}

@article{pidgaykoVoltagetunableOpticalParametric2023,
  title = {Voltage-Tunable Optical Parametric Oscillator with an Alternating Dispersion Dimer Integrated on a Chip},
  author = {Pidgayko, Dmitry and Tusnin, Aleksandr and Riemensberger, Johann and Stroganov, Anton and Tikan, Alexey and Kippenberg, Tobias J.},
  year = 2023,
  month = nov,
  journal = {Optica},
  volume = {10},
  number = {11},
  pages = {1582},
  issn = {2334-2536},
  doi = {10.1364/OPTICA.503022},
  urldate = {2025-08-17},
  langid = {english}
}

@article{qiLowlossHighlyTunable2025,
  title = {Low-Loss, Highly Tunable {{Sagnac}} Loop Reflectors and {{Fabry}}--{{P\'erot}} Cavities on Thin-Film Lithium Niobate},
  author = {Qi, Luke and Khalatpour, Ali and Herrmann, Jason F. and Park, Taewon and Dean, Devin and Robison, Sam and Hwang, Alexander and Stokowski, Hubert and Serkland, Darwin and Fejer, Martin M. and {Safavi-Naeini}, Amir H.},
  year = 2025,
  month = aug,
  journal = {Optics Letters},
  volume = {50},
  number = {16},
  pages = {5173},
  issn = {0146-9592, 1539-4794},
  doi = {10.1364/OL.568165},
  urldate = {2025-09-26},
  langid = {english}
}

@article{shimTunableSinglemodeChipscale2021,
  title = {Tunable Single-Mode Chip-Scale Mid-Infrared Laser},
  author = {Shim, Euijae and {Gil-Molina}, Andres and Westreich, Ohad and Dikmelik, Yamac and Lascola, Kevin and Gaeta, Alexander L. and Lipson, Michal},
  year = 2021,
  month = dec,
  journal = {Communications Physics},
  volume = {4},
  number = {1},
  pages = {268},
  issn = {2399-3650},
  doi = {10.1038/s42005-021-00770-6},
  urldate = {2025-08-17},
  langid = {english}
}

@article{siaAnalysisCompactSilicon2020,
  title = {Analysis of {{Compact Silicon Photonic Hybrid Ring External Cavity}} ({{SHREC}}) {{Wavelength-Tunable Laser Diodes Operating From}} 1881--1947 Nm},
  author = {Sia, Jia Xu Brian and Wang, Wanjun and Qiao, Zhongliang and Li, Xiang and Guo, Tina Xin and Zhou, Jin and Littlejohns, Callum G. and Liu, Chongyang and Reed, Graham T. and Wang, Hong},
  year = 2020,
  month = dec,
  journal = {IEEE Journal of Quantum Electronics},
  volume = {56},
  number = {6},
  pages = {1--11},
  issn = {0018-9197, 1558-1713},
  doi = {10.1109/JQE.2020.3029964},
  urldate = {2025-08-17},
  copyright = {https://ieeexplore.ieee.org/Xplorehelp/downloads/license-information/IEEE.html}
}

@article{siddharthUltrafastTunablePhotonicintegrated2025,
  title = {Ultrafast Tunable Photonic-Integrated Extended-{{DBR Pockels}} Laser},
  author = {Siddharth, Anat and Bianconi, Simone and Wang, Rui Ning and Qiu, Zheru and Voloshin, Andrey S. and Bereyhi, Mohammad J. and Riemensberger, Johann and Kippenberg, Tobias J.},
  year = 2025,
  month = jul,
  journal = {Nature Photonics},
  volume = {19},
  number = {7},
  pages = {709--717},
  issn = {1749-4885, 1749-4893},
  doi = {10.1038/s41566-025-01687-0},
  urldate = {2025-09-26},
  langid = {english}
}

@article{taschlerFemtosecondPulsesMidinfrared2021,
  title = {Femtosecond Pulses from a Mid-Infrared Quantum Cascade Laser},
  author = {T{\"a}schler, Philipp and Bertrand, Mathieu and Schneider, Barbara and Singleton, Matthew and Jouy, Pierre and Kapsalidis, Filippos and Beck, Mattias and Faist, J{\'e}r{\^o}me},
  year = 2021,
  month = dec,
  journal = {Nature Photonics},
  volume = {15},
  number = {12},
  pages = {919--924},
  issn = {1749-4885, 1749-4893},
  doi = {10.1038/s41566-021-00894-9},
  urldate = {2025-09-25},
  langid = {english}
}

@article{tranExtendingSpectrumFully2022a,
  title = {Extending the Spectrum of Fully Integrated Photonics to Submicrometre Wavelengths},
  author = {Tran, Minh A. and Zhang, Chong and Morin, Theodore J. and Chang, Lin and Barik, Sabyasachi and Yuan, Zhiquan and Lee, Woonghee and Kim, Glenn and Malik, Aditya and Zhang, Zeyu and Guo, Joel and Wang, Heming and Shen, Boqiang and Wu, Lue and Vahala, Kerry and Bowers, John E. and Park, Hyundai and Komljenovic, Tin},
  year = 2022,
  month = oct,
  journal = {Nature},
  volume = {610},
  number = {7930},
  pages = {54--60},
  issn = {0028-0836, 1476-4687},
  doi = {10.1038/s41586-022-05119-9},
  urldate = {2025-08-17},
  langid = {english}
}

@article{vainioSinglyResonantCw2008a,
  title = {Singly Resonant Cw {{OPO}} with Simple Wavelength Tuning},
  author = {Vainio, Markku and Peltola, Jari and Persijn, Stefan and Harren, Frans J. M. and Halonen, Lauri},
  year = 2008,
  month = jul,
  journal = {Optics Express},
  volume = {16},
  number = {15},
  pages = {11141},
  issn = {1094-4087},
  doi = {10.1364/OE.16.011141},
  urldate = {2025-09-26},
  langid = {english}
}

@article{wangWidelyTunableSinglemode2024,
  title = {Widely Tunable Single-Mode Interband Cascade Lasers Based on {{V-coupled}} Cavities and Dependence on Design Parameters},
  author = {Wang, Zhanyi and Gong, Jingli and He, Jian-Jun and Li, Lu and Yang, Rui Q. and Gupta, James A.},
  year = 2024,
  month = mar,
  journal = {Journal of Vacuum Science \& Technology B},
  volume = {42},
  number = {2},
  pages = {022204},
  issn = {2166-2746, 2166-2754},
  doi = {10.1116/6.0003376},
  urldate = {2025-08-17},
  langid = {english}
}

@article{weiUltrabroadbandTunableGaSbsilicon2025,
  title = {Ultra-Broadband Tunable {{GaSb-silicon}} Hybrid Laser for Gas Spectroscopy},
  author = {Wei, Jincheng and Lin, Songjian and Geng, Zhengqi and Yu, Hongguang and Chen, Yihang and Yang, Chengao and Niu, Zhichuan and Yu, Ying and Wang, Ruijun and Yu, Siyuan},
  year = 2025,
  month = oct,
  journal = {Photonics Research},
  volume = {13},
  number = {10},
  pages = {2913},
  issn = {2327-9125},
  doi = {10.1364/PRJ.568579},
  urldate = {2025-10-16},
  langid = {english}
}

@article{weiWidelyTunableShortwave2025,
  title = {Widely Tunable Short-Wave Mid-Infrared Hybrid Lasers Enabled by a Single Ultra-Compact Silicon Microring Resonator},
  author = {Wei, Jincheng and Geng, Zhengqi and Huang, Kan and Chen, Yihang and Yu, Ying and Yang, Chengao and Niu, Zhichuan and Wang, Ruijun and Yu, Siyuan},
  year = 2025,
  month = jul,
  journal = {Applied Physics Letters},
  volume = {127},
  number = {3},
  pages = {033301},
  issn = {0003-6951, 1077-3118},
  doi = {10.1063/5.0275617},
  urldate = {2025-10-16},
  langid = {english}
}

@article{xieHighefficiencyTunableLasers2025,
  title = {High-Efficiency Tunable Lasers Hybrid-Integrated with Silicon Photonics at 2.0 {\textbf{{$\mu$}}} m},
  author = {Xie, Yuxuan and McDonald, Corey A. and Morin, Theodore J. and Zhou, Zhican and Peters, Jonathan and Bowers, John E. and Wan, Yating},
  year = 2025,
  month = mar,
  journal = {Photonics Research},
  volume = {13},
  number = {3},
  pages = {737},
  issn = {2327-9125},
  doi = {10.1364/PRJ.550770},
  urldate = {2025-08-17},
  langid = {english}
}

@article{yangMidinfraredWidelyTunable2020,
  title = {Mid-Infrared Widely Tunable Single-Mode Interband Cascade Lasers Based on {{V-coupled}} Cavities},
  author = {Yang, Hanting and Yang, Rui Q. and Gong, Jingli and He, Jian-Jun},
  year = 2020,
  month = may,
  journal = {Optics Letters},
  volume = {45},
  number = {10},
  pages = {2700},
  issn = {0146-9592, 1539-4794},
  doi = {10.1364/OL.391308},
  urldate = {2025-08-17},
  langid = {english}
}

@article{ycasHighcoherenceMidinfraredDualcomb2018a,
  title = {High-Coherence Mid-Infrared Dual-Comb Spectroscopy Spanning 2.6 to 5.2 {$\mu$}m},
  author = {Ycas, Gabriel and Giorgetta, Fabrizio R. and Baumann, Esther and Coddington, Ian and Herman, Daniel and Diddams, Scott A. and Newbury, Nathan R.},
  year = 2018,
  month = apr,
  journal = {Nature Photonics},
  volume = {12},
  number = {4},
  pages = {202--208},
  issn = {1749-4885, 1749-4893},
  doi = {10.1038/s41566-018-0114-7},
  urldate = {2026-02-09},
  langid = {english}
}

@article{yuIntegratedFemtosecondPulse2022a,
  title = {Integrated Femtosecond Pulse Generator on Thin-Film Lithium Niobate},
  author = {Yu, Mengjie and Barton Iii, David and Cheng, Rebecca and Reimer, Christian and Kharel, Prashanta and He, Lingyan and Shao, Linbo and Zhu, Di and Hu, Yaowen and Grant, Hannah R. and Johansson, Leif and Okawachi, Yoshitomo and Gaeta, Alexander L. and Zhang, Mian and Lon{\v c}ar, Marko},
  year = 2022,
  month = dec,
  journal = {Nature},
  volume = {612},
  number = {7939},
  pages = {252--258},
  issn = {0028-0836, 1476-4687},
  doi = {10.1038/s41586-022-05345-1},
  urldate = {2025-09-26},
  langid = {english}
}

@article{yuSiliconchipbasedMidinfraredDualcomb2018a,
  title = {Silicon-Chip-Based Mid-Infrared Dual-Comb Spectroscopy},
  author = {Yu, Mengjie and Okawachi, Yoshitomo and Griffith, Austin G. and Picqu{\'e}, Nathalie and Lipson, Michal and Gaeta, Alexander L.},
  year = 2018,
  month = may,
  journal = {Nature Communications},
  volume = {9},
  number = {1},
  pages = {1869},
  issn = {2041-1723},
  doi = {10.1038/s41467-018-04350-1},
  urldate = {2026-02-09},
  langid = {english}
}

@article{zhangUltrabroadbandIntegratedElectrooptic2025,
  title = {Ultrabroadband Integrated Electro-Optic Frequency Comb in Lithium Tantalate},
  author = {Zhang, Junyin and Wang, Chengli and Denney, Connor and Riemensberger, Johann and Lihachev, Grigory and Hu, Jianqi and Kao, Wil and Bl{\'e}sin, Terence and Kuznetsov, Nikolai and Li, Zihan and Churaev, Mikhail and Ou, Xin and {Santamaria-Botello}, Gabriel and Kippenberg, Tobias J.},
  year = 2025,
  month = jan,
  journal = {Nature},
  volume = {637},
  number = {8048},
  pages = {1096--1103},
  issn = {0028-0836, 1476-4687},
  doi = {10.1038/s41586-024-08354-4},
  urldate = {2025-09-26},
  langid = {english}
}

@article{zhouHighPerformanceMonolithic2017,
  title = {High Performance Monolithic, Broadly Tunable Mid-Infrared Quantum Cascade Lasers},
  author = {Zhou, Wenjia and Wu, Donghai and McClintock, Ryan and Slivken, Steven and Razeghi, Manijeh},
  year = 2017,
  month = oct,
  journal = {Optica},
  volume = {4},
  number = {10},
  pages = {1228},
  issn = {2334-2536},
  doi = {10.1364/OPTICA.4.001228},
  urldate = {2025-08-17},
  langid = {english}
}

@article{zhouProspectsApplicationsOnchip2023,
  title = {Prospects and Applications of On-Chip Lasers},
  author = {Zhou, Zhican and Ou, Xiangpeng and Fang, Yuetong and Alkhazraji, Emad and Xu, Renjing and Wan, Yating and Bowers, John E.},
  year = 2023,
  month = jan,
  journal = {eLight},
  volume = {3},
  number = {1},
  pages = {1},
  issn = {2662-8643},
  doi = {10.1186/s43593-022-00027-x},
  urldate = {2025-09-26},
  langid = {english}
}

@article{zhuIntegratedPhotonicsThinfilm2021a,
  title = {Integrated Photonics on Thin-Film Lithium Niobate},
  author = {Zhu, Di and Shao, Linbo and Yu, Mengjie and Cheng, Rebecca and Desiatov, Boris and Xin, C. J. and Hu, Yaowen and Holzgrafe, Jeffrey and Ghosh, Soumya and {Shams-Ansari}, Amirhassan and Puma, Eric and Sinclair, Neil and Reimer, Christian and Zhang, Mian and Lon{\v c}ar, Marko},
  year = 2021,
  month = jun,
  journal = {Advances in Optics and Photonics},
  volume = {13},
  number = {2},
  pages = {242},
  issn = {1943-8206},
  doi = {10.1364/AOP.411024},
  urldate = {2025-09-26},
  langid = {english}
}

@article{zouHighcapacityFreespaceOptical2022,
  title = {High-Capacity Free-Space Optical Communications Using Wavelength- and Mode-Division-Multiplexing in the Mid-Infrared Region},
  author = {Zou, Kaiheng and Pang, Kai and Song, Hao and Fan, Jintao and Zhao, Zhe and Song, Haoqian and Zhang, Runzhou and Zhou, Huibin and Minoofar, Amir and Liu, Cong and Su, Xinzhou and Hu, Nanzhe and McClung, Andrew and Torfeh, Mahsa and Arbabi, Amir and Tur, Moshe and Willner, Alan E.},
  year = 2022,
  month = dec,
  journal = {Nature Communications},
  volume = {13},
  number = {1},
  pages = {7662},
  issn = {2041-1723},
  doi = {10.1038/s41467-022-35327-w},
  urldate = {2025-09-25},
  langid = {english}
}

\section{Methods}
\renewcommand{\figurename}{Supplementary FIG.}
\renewcommand{\thefigure}{\arabic{figure}}
\setcounter{figure}{0}

\begin{figure*}[t]
  \begin{center}
      \includegraphics[width=\textwidth]{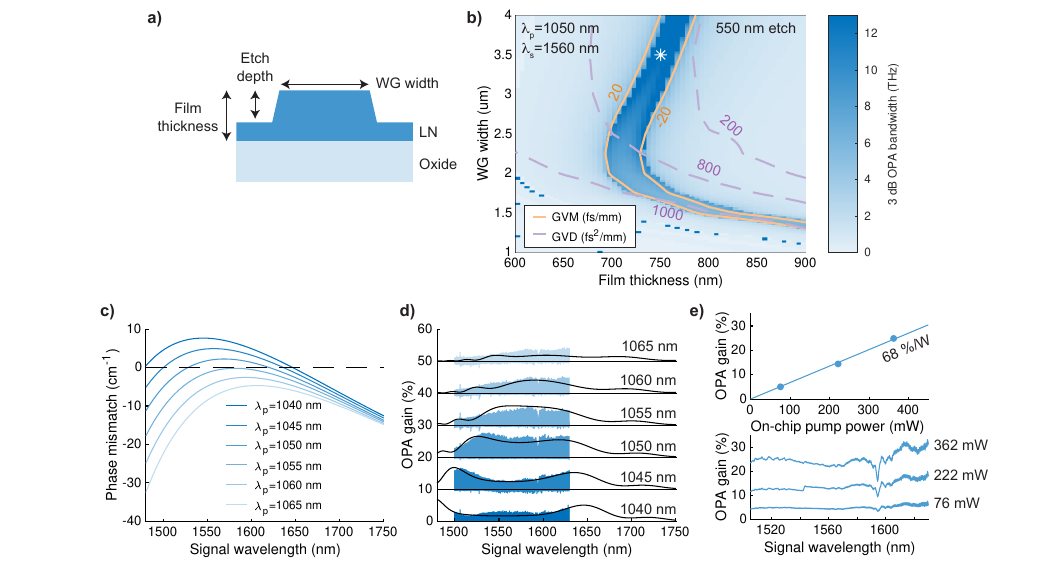}
  \end{center}
 \caption{\textbf{Broadband parametric gain via dispersion engineering.} a) Waveguide cross-section diagram with parameters used to parametrize the geometrical design search. b) OPA gain bandwidth as a 2D function of film thickness and waveguide width, for a fixed etch depth, pump wavelength, and signal center wavelength. Orange lines show the contours of signal-idler group velocity mismatch (GVM), and purple lines show the contours of the sum of signal and idler group-velocity dispersion (GVD). Both the GVM and GVD should be minimized to support the largest bandwidth. The white star symbol denotes the chosen design geometry ($3500$ nm WG width, $750$ nm film thickness). c) Simulated phase mismatch versus signal wavelength for the design geometry waveguide at six different pump wavelengths. The phase mismatch curves are centered around zero because we include the effect of the periodically-poled grating. d) Measured and simulated signal gain spectra for six different pump wavelengths. To match the measurements to simulation, a small spatially-dependent quadratic film thickness variation is included. e) Bottom: OPA signal gain spectra measured at three different on-chip pump powers. Top: Extracted small-signal gain as a function of pump power.}
 \label{fig:opa}
\end{figure*}

\subsection{OPA design, simulation, and measurement}

We first search for a waveguide geometry that enables broad optical parametric gain. Three parameters define the LN-on-oxide waveguide geometry: film thickness, etch depth, and waveguide width (Fig. \ref{fig:opa}a). The etch angle ($12$~degrees) is known from scanning electron microscopy. For a given waveguide geometry, we consider a three-wave interaction centered at pump wavelength $\lambda_p=1050$~nm, signal $\lambda_s=1560$~nm, and idler $\lambda_i = \left(1/\lambda_p - 1/\lambda_s\right)^{-1}$. Around these wavelengths the phase mismatch can be expanded in terms of the signal-idler group-velocity mismatch ($\Delta k'_{si}=1/v_{g,s} - 1/v_{g,i}$) and summed signal/idler group-velocity dispersion ($k''_{si}=k''_s+k''_i$) \cite{jankowskiDispersionengineeredH2Nanophotonics2021b}:
\begin{equation}
    \Delta k(\Omega) = \Delta k'_{si} \Omega - \frac{1}{2} k_{si}'' \Omega^2
\end{equation}
where $\Omega$ ($-\Omega$) is the angular frequency deviation of the signal (idler) around its central wavelength for a fixed pump. The gain spectrum is (in the low gain limit): $\text{sinc}(\Delta k(\Omega)L_{s}/2)^2$, where $L_{s}$ is the length of the gain section, in our case $9$~mm. Clearly, minimizing both $\Delta k'_{si}$ and $k''_{si}$ increases the gain bandwidth. Fig. \ref{fig:opa}b plots the $3$-dB bandwidth of the gain spectrum for a fixed $550$~nm etch depth for different waveguide widths and film thicknesses. In addition, we plot the GVM contours ($\Delta k'_{si}$) and summed GVD contours ($k''_{si}$) that in our case are primarily dominated by idler GVD. As evidenced in the plot, geometries that produce the largest bandwidths primarily follow the contour of GVM, and minimizing the GVD along that contour with larger waveguide widths/film thicknesses maximizes the bandwidth further. Plots for different etch depths have similar features, requiring large film thicknesses (${>}650$~nm) and waveguide widths (${>}2$~\textmu m) for large bandwidths. We choose a deep etch depth ($550$~nm) to provide strong modal confinement at pump, signal, and idler wavelengths. We choose a film thickness ($750$~nm) and waveguide width ($3.5$~\textmu m) to provide broadband gain in accordance with the dispersion engineering. By choosing a geometry that eliminates GVM, the simulated fixed-pump phase mismatch $\Delta k(\lambda_s) = \Delta k_{WG}(\lambda_s) - 2\pi/\Lambda$ appears quadratic versus wavelength (Fig. \ref{fig:opa}c), where $\Delta k_{WG}(\lambda_s) = k_p - k_s(\lambda_s) - k_i(\lambda_i)$ and $\Lambda$ is the periodic poling period.

Next we take OPA gain measurements to verify the dispersion engineering. We send quasi-CW pump pulses to the periodically-poled gain waveguide using an acousto-optic modulator that chops light from our $1$~\textmu m laser into $20$~\textmu s pulses at $10$~kHz repetition rate. Meanwhile, we couple a signal seed laser (Santec TSL-570) into the same waveguide and scan over its full range ($1500$-$1630$~nm). We collect the signal light exiting the chip with a nanosecond photodetector (Newport 1623). By comparing the signal strength when the pump pulse is turned on versus off, we can directly extract the signal gain, plotted in Fig. \ref{fig:opa}d and Fig. \ref{fig:fig1}d. We also use a lock-in amplifier to process the collected signal to calibrate the on-chip pump power-dependent gain (Fig. \ref{fig:opa}e) of $68$~\%/W. When accounting for the ($1.7$x) reduction of peak gain from film thickness variations, this number is $4$x smaller than the simulations predict.

Finally, we validate the dispersion engineering by matching experimental transfer functions to simulation (Fig. \ref{fig:opa}d). Explaining the experimental data requires a small spatially-dependent phase mismatch term: $\Delta k_{tot}(z, \lambda_s) = \Delta k(\lambda_s) + \Delta \tilde{k}(z)$. Spatially-dependent phase mismatch typically arises from film thickness variations \cite{chenAdaptedPolingBreak2024} and broadens the ideal $\text{sinc}^2(\Delta k L_s/2)$ transfer function. The resultant transfer function is then calculated numerically by solving the coupled nonlinear equations:
\begin{equation}
\begin{cases}
      \partial_z a_i(z) = \gamma a_s^*(z) h(z) , \\
      \partial_z a_s^*(z) = \gamma^* a_i(z) h^*(z)\\
\end{cases} 
\end{equation}
where $\gamma = -i\sqrt{\eta_{DFG}} \sqrt{\omega_s/\omega_i} A_{p}(0)$, $\eta_{DFG}$ is the nonlinear DFG generation efficiency, $a_s$ ($a_i$) are the photon number normalized amplitudes of signal (idler), and $A_p$ is the power normalized pump amplitude. The phase integral $h$ is:
\begin{equation}
    h(z) = e^{-i\int_0^z \Delta k_{tot}(z')dz'}.
\end{equation}
We find that incorporating $2$~nm of total film thickness variation with slight quadratic spatial dependence matches the simulation to experiment (Fig. \ref{fig:opa}d). The result of including this spatial dependence broadens the gain spectrum 3 dB bandwidth from $13$ to $18$~THz and reduces the peak nonlinear efficiency by ~$40$\%. The pump wavelength and signal wavelength-dependent gain spectra now match experiment well (Fig. \ref{fig:opa}d). The double-peaked transfer function indicates successful realization of phase mismatch depending quadratically on detuning and points to a favorable range of pump wavelengths of $1045$-$1050$~nm to produce the broadest gain.

\subsection{Vernier cavity design and measurement}

\begin{figure*}[t]
  \begin{center}
      \includegraphics[width=\textwidth]{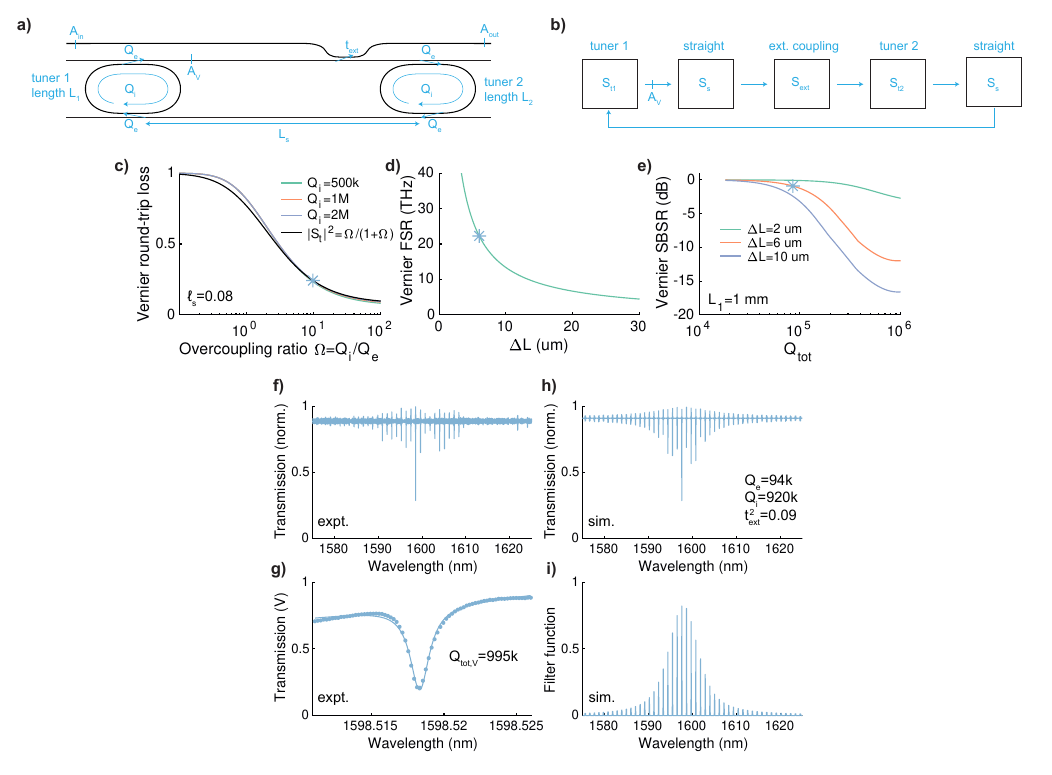}
  \end{center}
 \caption{\textbf{Vernier cavity design and measurement.} a) Diagram of the Vernier resonator, depicting important quantities used in the design/modeling. b) S-matrix model of the Vernier cavity, representing the concatenation of two tuner cavity elements, two straight sections, and an external coupler. c) Round-trip loss of the Vernier mode as a function of each tuner cavity's overcoupling ratio, and assuming a vertical offset of $8$\% loss from the straight sections. The result for the full S-matrix model for three different values of the tuner cavity intrinsic Q-factor $Q_i$ are shown alongside the result computed using an approximation of each tuner cavity's transfer function. d) Vernier free-spectral range as a function of tuner cavity length offset $\Delta L$. e) Vernier sideband suppression ratio as a function of each tuner cavity's total Q-factor $Q_{tot}$, for three different values of $\Delta L$. The blue star in (c-e) shows where the experimentally estimated device parameters. f) Measured Vernier cavity transmission as a function of signal wavelength, taken from a tunable laser scan. g) Zoomed-in transmission of the main Vernier mode, fitted to a Lorentzian lineshape. h) Simulated Vernier cavity transmission for device parameters optimized to match experiment. i) Simulated Vernier filter transmission function, corresponding to the transmission scan of h).}
 \label{fig:vernier_design}
\end{figure*}

To incorporate the Vernier effect within our OPO, we employ a Vernier "racetrack" configuration with two double-sided tuner racetrack cavities (tuner 1 and 2) connected by straight sections (Fig.~\ref{fig:vernier_design}a). The Vernier racetrack configuration can pose difficulties in integrated lasers because of mode competition between clockwise/counter-clockwise modes, but this is not a problem in our OPO because the pump wave only amplifies co-propagating signals. 

The tuner cavities have lengths $L_1$ and $L_2$ with nominally identical intrinsic ($Q_i$) and extrinsic ($Q_e$) quality factors. Each of the tuner cavities' two feedlines contributes the extrinsic coupling $\kappa_e = \omega/Q_e$. The overcoupling ratio $\Omega=Q_i/Q_e$ plays an important role in the analysis. Straight waveguide connections have a length $L_s=9$~mm and are assumed to contribute a small propagation loss ($8$~\% total, corresponding to $Q_s=2$~million). An external waveguide coupler on one of the straight sections introduces a field coupling coefficient of $t_{ext}$. It is useful to describe the tuner cavities, the straight waveguides, and the external coupler with the field transfer functions: $S_{t1,2}$, $S_s$, and $S_{ext}$ (Fig.~\ref{fig:vernier_design}b). 

Round-trip Vernier loss can be computed by assuming both tuner cavities are on resonance and have a power transfer of 
\begin{equation}
    |S_{t1}|^2 = |S_{t2}|^2 = |S_{t}|^2 = \frac{t^4P}{(1-r^2P)^2}
\end{equation}
where $t=\sqrt{1-r^2}$ is the field coupling coefficient for each tuner cavity's external coupler, and $P$ is the round-trip field transmission in each tuner cavity. In the overcoupled limit, the transmission simplifies to $|S_{t}|^2\approx \Omega / (1 + \Omega)$, which describes the round-trip loss well for a range of reasonable values of $Q_i$ (Fig.~\ref{fig:vernier_design}c). Our levels of parametric gain require round-trip losses less than ${\sim}30$\%. Experimental devices achieved $\Omega \approx 10$, resulting in a round-trip loss near $25$~\%.

The Vernier tuning range is set by the difference in free spectral range between the two tuner cavities (Fig.~\ref{fig:vernier_design}d). The Vernier FSR is given by $FSR_V = c/(n_g|\Delta L|)$ where $\Delta L = L_1 - L_2$ is the length difference between the two cavities. To fully utilize the $20$~THz parametric gain bandwidth, we choose $\Delta L=6$~\textmu m.

Vernier sideband suppression ratio (SBSR) is defined here as the ratio between transmission at the Vernier main peak and the first side peak at one tuner cavity FSR away. SBSR depends on how the spectra overlap between each cavity's lineshape, and is therefore dependent on each cavity's total linewidth and the FSR mismatch. We simulate SBSR and its dependence on both $\Delta L$ and each tuner cavity's total Q-factor $Q_{tot} = (1/Q_i + 2/Q_e)^{-1}$, for the utilized value of $L_{1}=1$~mm. Results in Fig.~\ref{fig:vernier_design}e indicate a SBSR of approximately 1~dB in the chosen design.

For device characterization, light from a tunable telecom laser (Santec TSL-570) couples into the external bus waveguide, which feeds into the Vernier cavity. Raw transmission spectra are cleaned and normalized by dividing with the low-pass filtered signal. Tuning the Vernier cavity near $1600$~nm reveals distinct Vernier modes (Fig. \ref{fig:vernier_design}f) with sidebands suppression of around 3~dB. The principal Vernier peak exhibits total Q-factor $Q_{tot,V}\sim 995\times 10^3$. To fit the experimental data, we use the transfer matrix formalism to model steady state transmission through the external bus waveguide: $T_{ex} = |-t_{ext} + \eta_{ext}S_{t2}S_sS_{t1}S_s|^2$. By matching parameters to experiment, we extract $Q_e=94\times 10^3$, $Q_i = 920 \times 10^3$, and $t_{ext}^2=9$~\%. Simulated spectra closely match experimental SBSR and $Q_{tot,V}$. The simulated round-trip power transfer function (Fig.~\ref{fig:vernier_design}f) displays smaller sideband suppression than the experimental transmission plot, which benefits from incomplete alignment between Vernier longitudinal modes and the filter transfer function, slightly enhancing observed SBSR.

\subsection{Device fabrication}

We designed a fabrication flow that can be compatible with future wafer-scale manufacturing.

First, we prepare the chip for waveguides fabrication with thinning and periodic poling. We begin with a 12x14mm $900$~nm TFLN-on-oxide die (NanoLN), then thin it to $750$~nm film thickness using an argon ion mill (IntlVac). Next, we pattern poling electrodes using photolithography (Heidelberg Instruments) on a LOR/SPR bilayer resist stack. We deposit the poling electrode metal, Al, with electron beam evaporation, then liftoff the resist in heated N-methyl-pyrrolidone. We pole the devices by applying ${\sim}1$~kV, ${\sim}20$~ms electrical pulses, verifying the domain growth with a home-built second harmonic generation microscope. After poling is complete, we clean the electrodes off with weak tetramethyl ammonium hydroxide (MF-319). 

Next, we etch waveguides. First, we grow $1$~\textmu m of silica by PECVD to serve as an etch hardmask. Then we pattern the hardmask using electron-beam lithography on a MaN-2410 resist (which is also photolithography compatible) followed by reactive ion etching with fluorine chemistry. We clean the MaN softmask using ozone treatment and piranha solution. Afterwards, we etch the LN waveguides by $550$~nm in an ion mill and clean the waveguides with hydrofluoric acid and piranha solution.

We then add heater electrodes to the devices. The heater electrodes ($10$ nm/$40$ nm Ti/Pt stack) are deposited directly on lithium niobate using the same photolithography process as for the poling electrodes described above. Typical electrical resistances are $1.3 \textrm{ k}\Omega$ for the tuner cavity electrodes and $20 \textrm{ k}\Omega$ for the phase shifter electrode.

Finally, we use laser stealth dicing (DISCO Corporation) to define clean waveguide facets to couple light in/out of the chip.

\subsection{OPO optical spectra measurements}

Our $1$~\textmu m pump light originates from a tunable external cavity diode laser (Toptica DL Pro) that we park at a fixed wavelength. Then, the light is amplified by a Ytterbium-doped fiber amplifier (Civil Laser). To create quasi-CW pump pulses, the light is chopped by an acousto-optic modulator at $10$~kHz repetition rate, with $2$\% duty cycle (Figs.~\ref{fig:fig2},~\ref{fig:fig3}) or $5$\% duty cycle (Fig.~\ref{fig:fig4}). After passing through a manual polarization controller to set TE light delivery, the light couples to the chip through a single-mode Hi1060 lensed fiber (OZ Optics). We calibrate the fiber-to-chip coupling to be $18.5 \pm 1$\%.

The OPO light generated from the chip is collected in two ways: (1) with a zinc fluoride glass multimode fiber (La Verre Fluore) that supports signal/idler wavelengths or (2) with a Hi1060 single-mode fiber that only supports signal wavelengths. We use the multimode fiber (Fig. \ref{fig:fig2}b,c,e,f; Fig. \ref{fig:fig3}b,d) to measure both near- and mid-infrared optical spectra using a Yokogawa AQ6376 optical spectrum analyzer (OSA). Only $0.4$\% of off-chip mid-infrared light is collected by the OSA, and using a large-core multimode fiber significantly reduces the wavelength resolution on the OSA due to production of a multimode speckle pattern. For broad tuning plots (Fig. \ref{fig:fig2}b,e) we optimize for scan speed and use low wavelength resolution ($2$~nm), while for narrow tuning plots (Fig. \ref{fig:fig2}c, \ref{fig:fig3}b,d) we use a higher resolution setting ($200$~pm for signal, $1$~nm for idler). We use the single-mode fiber (Fig. \ref{fig:fig2}d, \ref{fig:fig3}d) to obtain the highest resolution ($20$~pm) spectra of the signal wave available to us.

The chip is controlled at fixed global temperature using a thermo-electric cooler (Thorlabs TECD2) to arrive at the correct phase matching wavelengths. Device 1 (Figs.~\ref{fig:fig1}, \ref{fig:fig2}, \ref{fig:fig3}d, \ref{fig:fig4}, \ref{fig:fig5}) is held at $180$~C, and Device 2 (Figs.~\ref{fig:fig3}a, \ref{fig:fig5}) is held at $120$~C. The operating temperatures can be lowered by adjusting the poling period in a future run. To tune the OPO output, integrated phase shifters are controlled using DC voltage signals.

\subsection{OPO power characterization}

For power characterization (Fig.~\ref{fig:fig4}), we use the same input light delivery as in the OPO spectral measurements, described above. On the output side, telecom signal light is again collected with the single-mode lensed fiber, then sent into a photodetector (Newport 1623). We again collect mid-infrared idler light with the zinc fluoride glass multimode fiber, then focus it in free space onto a MCT photodetector (Thorlabs PDAVJ5). An important part of this measurement is calibrating the relationship between the mid-infrared light off-chip to the photodetected voltage signal. To do so, we use CW difference frequency generation on a neighboring straight poled waveguide, which is directly detected off-chip (${\sim}0.8$~mW) with a thermal power sensor (Thorlabs S401C). Then, we use the output chain to measure the light (multimode fiber$\rightarrow$collimator$\rightarrow$beamsplitter$\rightarrow$lens$\rightarrow$detector), which allows us to extract a $0.66\pm0.08$\% efficiency from off-chip to detector. 

\subsection{Effect of modal hybridization}

\begin{figure*}[t]
  \begin{center}
      \includegraphics[width=\textwidth]{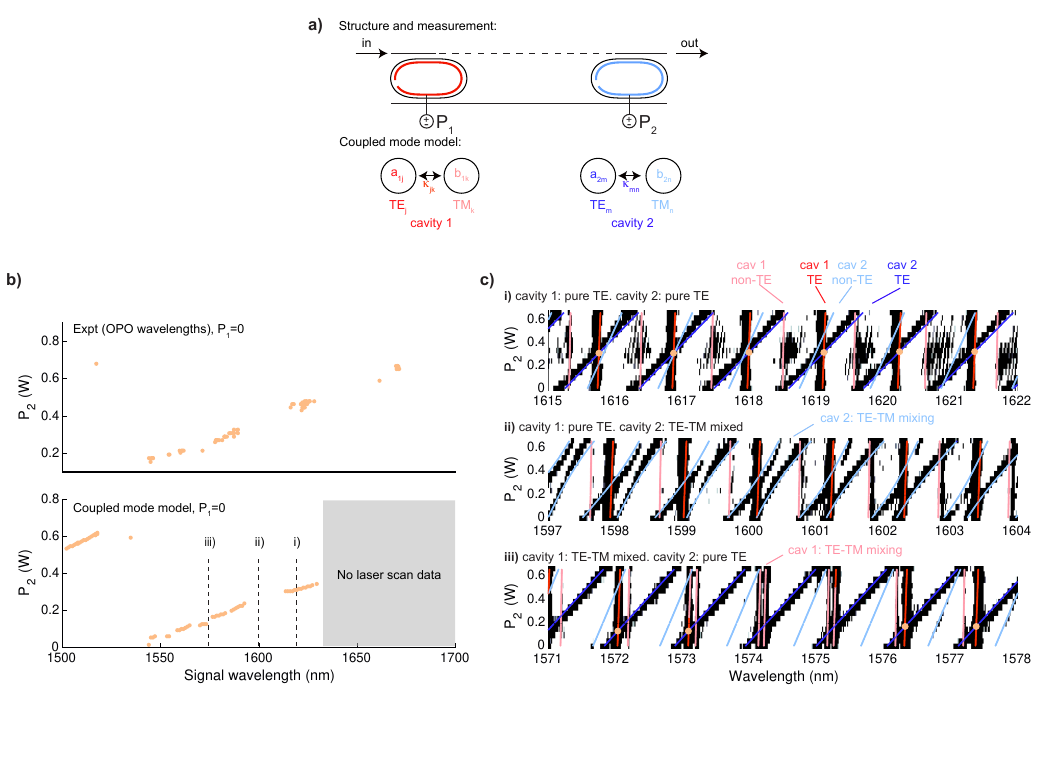}
  \end{center}
 \caption{\textbf{Mode-crossing-limited OPO tuning.} a) Diagram depicting the coupled-mode model we use to describe the OPO tuning behavior. We model each tuner cavity having coupling between TE and TM resonant modes, and do not consider coupling between the tuner cavities. b) Top: Measured OPO signal output wavelengths as a function of heater 2 power, for a fixed $P_1=0$. Bottom: Simulated Vernier mode positions as a function of heater 2 power. c) Colorplot: heater power-dependent laser transmission wavelength scans. Red/blue lines: Simulated eigenmode frequencies of tuner cavity 1/2. Dark red/blue lines indicate the eigenmode from cavity 1/2 has ${>}90$\% TE content, and lighter red/blue lines indicate ${<}90$\% TE content. When ${<}90$\% TE modes intersect, we include an orange point representing the expected OPO wavelength from the coupled mode model. We highlight three wavelength regions: i) clean, pure-TE modes from both cavity 1 and cavity 2; ii) clean, pure-TE modes from cavity 1 with hybridized modes from cavity 2; iii) hybridized modes from cavity 1 with clean, pure-TE modes from cavity 2.}
 \label{fig:mode_x}
\end{figure*}

To understand why OPO tuning gaps arise, we build a coupled mode model to describe the TE/TM modal hybridization we observe in the passive spectra of the tuner cavities (Fig.~\ref{fig:mode_x}). The model (Fig.~\ref{fig:mode_x}a) treats each tuner cavity independently without assuming any cross-coupling between tuner cavities. Cavity 1's eigenmode frequencies are computed by first assuming a set of unperturbed TE/TM basis modes ($a_{1j}$/$b_{1k}$) and applying a coupling coefficient $\kappa_{jk}$ that mixes TE with TM. The same process is used for cavity 2. The model parameters are tuned to match the cavity 1/2's eigenmode frequencies to experimental data from wavelength-scanning a telecom laser on the straight waveguide that couples to both tuner cavities (Fig.~\ref{fig:mode_x}a). 

The OPO oscillation wavelength is simulated by selecting the wavelength where the eigenfrequencies of cavity 1/2 align. This process is repeated for each integrated phase shifter setting to construct a whole simulated version of the OPO tuning curve (Fig.~\ref{fig:mode_x}b). Comparing to experimental OPO data shows that the spectral position and size of gaps are qualitatively well-produced using the coupled mode model. In addition, the model shows that even without tuning gaps, the OPO wavelengths can drift slightly away from a simple linear tuning curve.

Fig.~\ref{fig:mode_x}c gives a closer insight into what produces spectral gaps, based on the experimental laser scan data on the passive tuner cavity modes. The colormap plots the combined passive response of tuner cavities 1/2 versus wavelength, while $P_2$ is scanned from $0\rightarrow700$~mW. Dark areas in the colormap indicate the presence of a mode from either cavity 1 or 2. The computed eigenmodes from the coupled model are overlayed on the colormap. When eigenmodes from cavity 1/2 overlap and both have ${>}90$\% TE content, we mark the overlap location as a point on the simulated OPO tuning curve (orange dot). Fig.~\ref{fig:mode_x}c.i highlights a cleanly tuning region of OPO tuning curve near 1618~nm. Both cavity 1 and cavity 2's eigenmodes have strong TE content, and no mode hybridization is observed. Fig.~\ref{fig:mode_x}c.ii shows a portion of the tuning curve near 1600~nm where cavity 1's modes are clean TE, while cavity 2's modes have strong TE/TM hybridization. As a result, a gap in the OPO tuning curve is reported here. Fig.~\ref{fig:mode_x}c.iii shows the opposite situation, where cavity 2's modes are clean TE, while cavity 1's modes have TE/TM hybridization, which again causes a spectral gap.

\section*{Acknowledgments}
This work was supported by the Defense Advanced Research Projects Agency (DARPA) INSPIRED program, the National Science Foundation NSF-SNSF MOLINO project (No.~ECCS-2402483), and the Stanford School of Sustainability Accelerator. The authors wish to thank NTT Research for their financial and technical support. Device fabrication was performed at the Stanford Nano Shared Facilities (SNSF) and the Stanford Nanofabrication Facility (SNF), supported by NSF award ECCS-2026822. We gratefully acknowledge support from the Shoucheng Zhang Graduate Fellowship Program. This material is based upon work supported by the National Science Foundation Graduate Research Fellowship Program under Grant No.~DGE-1656518.

\section*{Author contributions}
A.Y.H., H.S.S., M.M.F., and A.H.S.-N. designed the device.
A.Y.H., H.S.S., and D.K.C. fabricated the device.
A.Y.H., H.S.S., L.Q., D.K.C., D.J.D., and T.P. developed fabrication procedures together. 
A.Y.H., H.S.S., G.H.A., and E.R. measured the device.
A.Y.H. and H.S.S. analyzed the data.
M.M.F. and A.H.S.-N. advised the project and provided experimental/theoretical support.
A.Y.H. drafted the manuscript with input from all the authors.

\section*{Competing interests}
H.S.S. and A.H.S.-N. are involved in developing lithium niobate technologies at ely Sensor Technologies, Inc. The remaining authors declare no competing interests.

\section*{Materials and Correspondence}
Material requests and correspondence should be addressed to A.H.S.-N. at safavi@stanford.edu.


\end{document}